\documentclass{aa}

\usepackage{graphicx}

\usepackage{txfonts}
\usepackage{bm}

\usepackage{xcolor}

\begin{document}

   \title{High-resolution models of the vertical shear instability}

   \author{G. Lesur
          \inst{1}
          \and
          H. Latter\inst{2}
          \and
          G.~I.~Ogilvie\inst{2}
          }

   \institute{Univ. Grenoble Alpes, CNRS, IPAG, 38000 Grenoble, France\\
              \email{geoffroy.lesur@univ-grenoble-alpes.fr}
         \and
             Department of Applied Mathematics and Theoretical Physics, Centre for Mathematical Sciences, University of Cambridge, Wilberforce Road, Cambridge CB3 0WA, UK\\
             \
             }

   \date{Received September 15, 1996; accepted March 16, 1997}

  \abstract
   {The vertical shear instability (VSI) is a promising mechanism for generating turbulence and transporting angular momentum in magnetically decoupled regions of protoplanetary discs. While most recent work has focused on adding more complex physics, the saturation properties of the instability in radially extended discs, and its convergence as a function of resolution, are still largely unknown.}
   {We address the question of VSI saturation and associated turbulence using radially extended, fully 3D global disc models with very high resolution in the locally isothermal approximation, to capture both the largest VSI scales and the small-scale turbulent cascade. }
   {We used the GPU-accelerated code Idefix to achieve resolutions of up to 200 points per scale height in the three spatial directions. We chose numerical techniques that minimise numerical diffusion as much as possible: third-order reconstruction schemes, orbital advection, and a third-order time integrator.  We modelled the VSI in disc domains extending up to $R_\mathrm{out}/R_\mathrm{in}=7$ in the highest resolution case and $R_\mathrm{out}/R_\mathrm{in}=25$ in our intermediate model (100 points per scale height), with a full $2\pi$ azimuthal extent and a disc aspect ratio $H/R=0.1$. }
   {We demonstrate that large-scale transport properties converge with 100 points per scale height, leading to a Shakura-Sunyaev $\alpha=1.3\times 10^{-3}$ in the bulk of the computational domain. Inner boundary condition artefacts propagate deep inside the computational domain (typically over $\Delta R/R\sim 2-3$), leading to a reduced $\alpha$ in these regions. The large-scale corrugation wave zones identified in 2D models persist in 3D, albeit with less coherence. Our models show no signs of long-lived zonal flows, pressure bumps, or vortices, in contrast to lower-resolution simulations. Finally, we show that the turbulent cascade resulting from VSI saturation can be interpreted within the framework of critically balanced rotating turbulence. We propose that the small non-axisymmetric scales could be modelled with an effective anisotropic viscosity in 2D simulations, significantly reducing the computational cost of these models while still capturing the important physics. }
   {The VSI leads to vigorous turbulence in protoplanetary discs, associated with outward angular momentum transport, but without any significant long-lived features that could enhance planet formation. The innermost regions of VSI simulations are consistently polluted by boundary-condition artefacts influencing the first VSI wave train. Radially extended domains should therefore be used in a more systematic manner, and more realistic inner boundaries should be explored to mimic the radial structure of real protoplanetary discs. }


   \maketitle
%

\section{Introduction}

The vertical shear instability (VSI) has emerged as a prominent hydrodynamic mechanism capable of generating turbulence in regions of protoplanetary discs where magnetic activity is weak or absent, a region usually referred to as the dead zone. First identified in the early work of \cite{Urpin.Brandenburg98}, the VSI gained substantial attention following the seminal study by \cite{Nelson.Gressel.ea13}, which demonstrated its robust development in locally isothermal discs. The instability is now recognised as a manifestation of the Goldreich-Schubert-Fricke (GSF) instability, originally formulated in the context of stellar interiors \citep{Goldreich.Schubert67,Fricke68}, and is driven by vertical shear in the angular velocity profile under conditions of efficient cooling.

Since then, linear analyses have clarified the conditions under which the VSI can grow, with key contributions from \cite{Barker.Latter15} and \cite{Lin.Youdin15}. The nonlinear saturation of the VSI has been explicitly addressed through 2D \citep{Flores-Rivera.Flock.ea20} and 3D \citep{Manger.Klahr.ea20,Shariff.Umurhan24} simulations and is believed to result either from delayed Kelvin–Helmholtz instabilities \citep{Latter.Papaloizou18} or from parametric instabilities \citep{Cui.Latter22}.

Most recent work on VSI has focused on adding new physics. For example, studies have examined the interaction of VSI with magnetohydrodynamics (MHD) and with the magneto-rotational instability \citep{Latter.Papaloizou18,Cui.Bai20,Latter.Kunz22,Cui.Bai22a}; thermodynamic improvements, such as the inclusion of radiative transfer with M1 closure schemes \citep{MelonFuksman.Flock.ea24}, and more realistic vertical disc structures \citep{Zhang.Zhu.ea24}; and finally, the impact of VSI on dust grain dynamics \citep{Flock.Nelson.ea17,Dullemond.Ziampras.ea22,Pfeil.Birnstiel.ea23}, together with the feedback of the grains' inertia on the VSI \citep{Huang.Bai25}.

Despite this additional complexity, many basic questions remain regarding the long-term saturation and energy cascade of the VSI. Notably, one of the defining features of the VSI is its ability to excite a type of large-scale inertial wave, known as `corrugation modes', which can dominate the vertical and radial dynamics of the disc \citep{Nelson.Gressel.ea13,Stoll.Kley14}. These corrugation modes form wave trains that develop into well-defined radial wave zones in 2D simulations \citep{Svanberg.Cui.ea22}. However, capturing several of these wave trains is notoriously difficult, as this requires radially extended simulation domains. The only studies that clearly exhibit several wave trains are axisymmetric \citep{Stoll.Kley14,Svanberg.Cui.ea22}, with $r_\mathrm{out}/r_\mathrm{in}>10$, while all 3D simulations published to date have  $r_\mathrm{out}/r_\mathrm{in}<5$ (cf.~tab.~\ref{tab:previousWork}). 

\begin{table}
      \caption[]{\label{tab:previousWork}Previous studies and their models' key properties, including the number of dimensions $D$, radial aspect ratio, number of grid points $n$ per disc scale height $H$ and azimuthal extent $\Delta \phi$. }
      \centering
         \begin{tabular}{lcccc}
            \hline
            \noalign{\smallskip}
            Reference      &  nD & $R_\mathrm{out}/R_\mathrm{in}$ & $n \,\textrm{per}\,H$ & $\Delta \phi$ \\
            \noalign{\smallskip}
            \hline
            \noalign{\smallskip}
            \cite{Stoll.Kley14} & 2D & 25    &   12  & /   \\ 
            \cite{Stoll.Kley14} & 3D & 5    &   12  & $\pi/2$   \\   
            \cite{Richard.Nelson.ea16} & 3D &  1.5  & $20^*$ & $\pi/4$ \\
            \cite{Manger.Klahr18} & 3D & 4  & 18 & $2\pi$ \\
            \cite{Flores-Rivera.Flock.ea20} & 2D & 10 & 203 & / \\
            \cite{Svanberg.Cui.ea22} & 2D & 30 & 12 & /\\
             \cite{Shariff.Umurhan24} & 3D &  2 & 74 & $2\pi$\\  
             \cite{Huang.Bai25a} & 3D & 3    &   $240^\dagger$  & $2\pi$ \\
             \cite{Huang.Bai25} & 3D & 3     &   30 & $\pi$ \\

             \noalign{\smallskip}
            \hline
         \end{tabular}
         \raggedright
         *: The resolution is very anisotropic. Here, we quote the resolution along $z$.\\
         $\dagger$: This resolution is achieved with adaptive mesh refinement only in a very small domain centred on the disc midplane
\end{table}

This question of radial extent is important because VSI is an instability of inertial waves travelling radially outwards \citep{Ogilvie.Latter.ea25}. Consequently, the innermost travelling wave packet is necessarily affected by the inner boundary condition of the domain, a problem that has been largely overlooked in the community. However, there are hints that some boundary artefacts are present:  \cite{Shariff.Umurhan24} show that the turbulent transport parameter $\alpha$ increases linearly with radius in a box with $r_\mathrm{out}/r_\mathrm{in}=2$, a rather unexpected result if the VSI and its saturation were truly a radially local process. Thus, how the VSI saturates in a truly radially extended domain remains an open question. 

Other questions of interest include whether the saturation mechanism involves the formation of long-lived structures such as zonal flows and vortices, which might trap dust grains and trigger planet formation \citep{Richard.Nelson.ea16,Manger.Klahr18,Pfeil.Klahr21}. Finally, the nature of the turbulent cascade is also debated, as various studies report different phenomenologies \citep{Manger.Klahr18,Shariff.Umurhan24,MelonFuksman.Flock.ea24}, ranging from direct Kolmogorov cascades to quasi-2D inverse cascades.

The goal of this work is to address the problem of VSI saturation in a sufficiently large domain to capture several corrugation wave zones, while at the same time ensuring sufficient resolution to adequately describe the small-scale turbulent cascade, thus providing a complete picture of VSI saturation. The models also need to be 3D to capture the possible breakup of zonal flows into vortices. The requirement for several wave zones translates into $r_\mathrm{out}/r_\mathrm{in}\sim 20$, while adequately capturing VSI saturation requires about $100$ points per disc pressure scale height in 2D models \citep{Flores-Rivera.Flock.ea20}, a constraint that we will assume to be valid in 3D. Combining these constraints implies simulations with $10^{10}$ to $10^{11}$ cells (i.e. about 400 times more cells than the most recent 3D models of \citealt{Shariff.Umurhan24}), making optimised numerical techniques and GPU acceleration mandatory.

In the following sections, we first present the numerical techniques and setup used. We then present the main results of our models and finally discuss its implications.

\section{Methods}

\subsection{Physical setup}

The setup consists of a locally isothermal disc with density $\rho\propto R^{-3/2}$, or equivalently surface density $\Sigma\propto R^{-1/2}$, and an aspect ratio $H/R=0.1$ such that the sound speed is $c_s=0.1v_K$, where $v_K$ is the Keplerian velocity. To capture radially extended structures, we considered simulations with radially extended domains. We constructed two simulations with $r_\mathrm{out}/r_\mathrm{in}=25$ at resolutions of $40$ and $80$ points per scale height (runs LX and HX), and a higher-resolution simulation with a reduced radial extension $R_\mathrm{out}/R_\mathrm{in}=7$, but an extreme resolution of $200$ points per scale height (run EX). All of the simulations extended vertically over $\pm 4$ pressure scale heights ($\theta-\pi/2 \in [-0.4,+0.4]$) and over $2\pi$ in the azimuthal direction. We used a logarithmic spacing for the radial grid, while keeping a constant spacing in $\theta$ and $\phi$, which ensured that the cell size per $H$ remained homogeneous throughout the domain.

We designed the radial boundary conditions to avoid wave reflection and to limit boundary artefacts as much as possible. At the boundary itself, we enforced an outflow condition, where the tangential velocity is copied into the ghost cells, while the normal velocity is either copied (when directed outside of the computational domain) or set to zero. The density and pressure are extrapolated from the last active zone assuming hydrostatic equilibrium power laws (see below).  We supplemented the radial boundary condition with a radial wave-killing zone in which the density, pressure, and velocity are relaxed towards the equilibrium values on a characteristic timescale equal to the local orbital timescale. These wave-killing zones extend radially over two local pressure scale heights on both sides of the domain.

The meridional boundary conditions are stress-free: the flow is not allowed to escape the domain, and the tangential velocity components are copied from the last active cell in the ghost cells. The density and pressure are extrapolated from the last active cell, assuming hydrostatic equilibrium. Finally, the flow is assumed to be periodic in the azimuthal direction.

We started simulation LX from a hydrostatic equilibrium \citep{Nelson.Gressel.ea13} on top of which we added a low-level random noise $\delta v/V_K=10^{-4}$. We restarted the higher-resolution simulations HX and EX from a snapshot taken at $t=1000$ orbits after the beginning of run LX, interpolating the field values onto the higher-resolution grid. This procedure allowed us to minimise the long transient preceding VSI saturation observed in run LX for the first 250 orbits.

\begin{table*}
      \caption[]{Main properties of simulations presented in the paper}
      \centering
         \begin{tabular}{lclcc}
            \hline
            \noalign{\smallskip}
            Name      &  $n_r\times n_\theta \times n_\phi$ & $R_\mathrm{in}, R_\mathrm{out}$ & $n_r\times n_\theta \times n_\phi \,\textrm{per}\,H$& $T_\mathrm{final} \textrm{(orbits at R=1)}$ \\
            \noalign{\smallskip}
            \hline
            \noalign{\smallskip}
            LX & $1288\times 400 \times 2512$ & $[1.0,25.0]$ & $40\times 50\times 40$ & $1500$   \\
            HX & $2576\times 800 \times 5024$ & $[1.0,25.0]$ & $80\times 100\times 80$ & $985$   \\
            EX & $3872\times 2000 \times 12544$ & $[1.0,7.0]$ & $200\times 250\times 200$ & $900$  \\

            \noalign{\smallskip}
            \hline
         \end{tabular}
     
\end{table*}
  
\subsection{Numerical technique}
The overarching goal of our numerical setup is to limit numerical dissipation, in order to resolve the VSI's secondary instabilities and the resulting turbulent cascade as accurately as possible. We performed all simulations presented here using the Idefix code \citep{Lesur.Baghdadi.ea23} with a compact third-order reconstruction scheme \citep[LiMO3]{Cada.Torrilhon09} and the Harten-Lax-van Leer-Contact (HLLC) Riemann solver.  To limit numerical dissipation due to advection by Keplerian rotation, we used the orbital advection scheme (also known as `Fargo') implemented in Idefix in a fully conservative manner \citep{Mignone.Flock.ea12}, using a piecewise parabolic (PPM) method to reconstruct fractional advection. The system of equations is evolved using a third-order TVD Runge--Kutta scheme.

We ran the code on the LUMI-G pre-exascale supercomputer in Finland, using 4096 AMD Mi250X GPUs simultaneously for the two highest-resolution models.

\subsection{Units and diagnostics}
 For all models, we use the inner radius as the length unit, while the Keplerian velocity at the inner radius serves as the velocity unit. The azimuthal average of a 3D quantity $Q$ is denoted $\langle Q \rangle$,  the time average of a quantity is given by $\langle Q \rangle_t$ (taken over the entire simulation time unless otherwise specified), and the azimuthal and vertical average over three scale heights $H$ is denoted by $\overline{Q}$, defined through the meridional angle $\theta_\pm = \frac{\pi}{2}\pm\arcsin(3H/R)$: 
 \begin{align}
 \overline Q = \frac{1}{\theta_+-\theta_-}\int_{\theta_-}^{\theta_+}\langle Q\rangle\sin\theta\,\mathrm{d}\theta.	
 \end{align}

The azimuthal average defines the non-axisymmetric deviation for each field as
$\bm{v'}=\bm{v}-\langle \bm{v}\rangle$. Finally, the Keplerian velocity profile is assumed to be constant on cylinders, i.e. $ v_K(r,\theta)=\big(r\sin\theta\big)^{-1/2}$. This is used to define the deviation from Keplerian rotation, $\delta \bm{v}(r,\theta,\phi)=\bm{v}-v_K\,\bm{e}_\phi$.
 
\section{Results}
\subsection{Overview}
\begin{figure*}
   \centering
   \includegraphics[width=\linewidth]{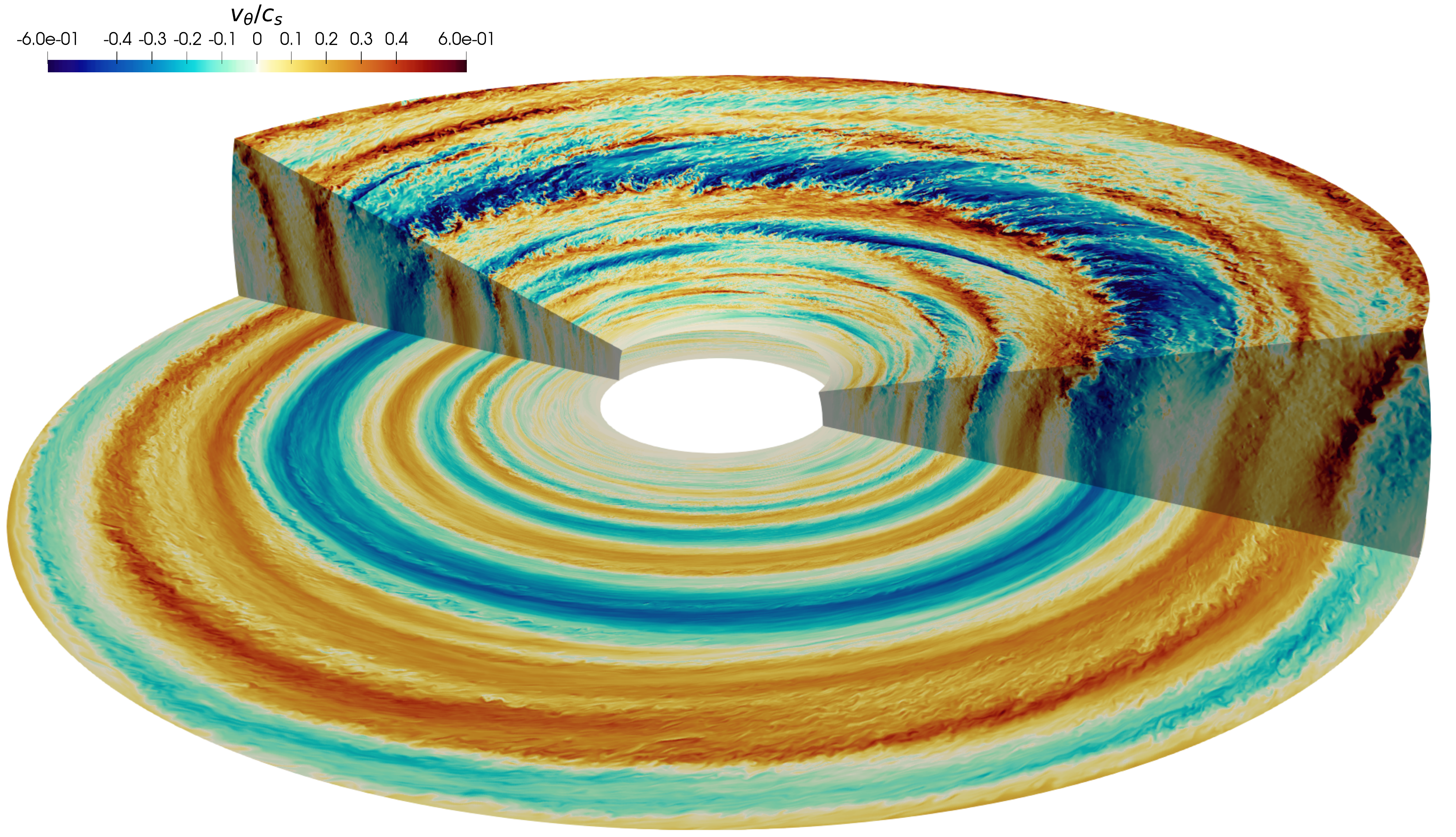}

   \caption{\label{fig:view3D}Three-dimensional rendering of the meridional velocity sonic Mach number $v_\theta/c_s$ in simulation EX (200 points per $H$) at $t=800$ orbits after restart. The front part shows the flow in the midplane, while the back part shows the flow at $3H$ above the midplane, with numerical boundaries at $4H$. The domain has been truncated at $R_\mathrm{out}=5.6$ to remove the outer wave-killing zone.}%
    \end{figure*}

The main features of our simulations can be identified in the instantaneous snapshot shown in Fig.~\ref{fig:view3D}, which shows the 3D latitudinal velocity $v_\theta$ of the highest-resolution model EX taken after 800 inner orbits. The flow exhibits large-scale, quasi-axisymmetric upward and downward motion visible as `bands'. In addition, the surface exhibits numerous fine-scale structures on top of these bands. We focus on each of these features in turn and demonstrate how they are connected.

\subsection{Time history and bulk transport properties}

\begin{figure}
   \centering
   \hspace{-5.0mm}\includegraphics[width=1.05\linewidth]{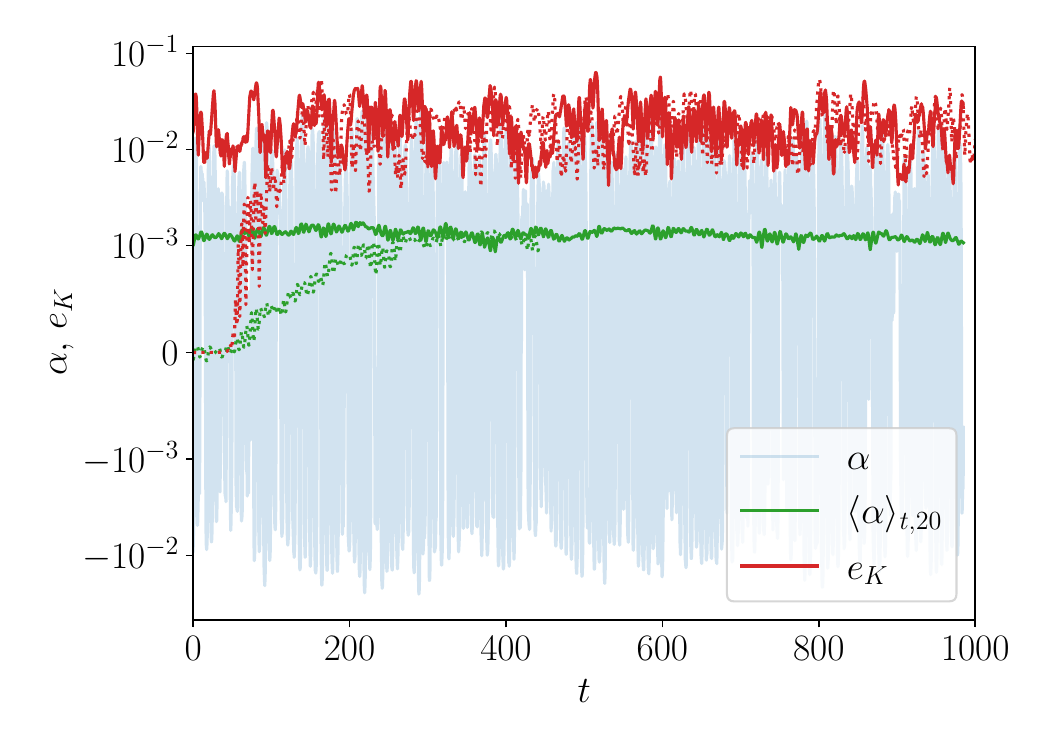}

   \caption{\label{fig:transport-t}Radial angular momentum coefficient ($\alpha$) and poloidal kinetic energy ($e_k$) measured at $R=4$ in run HX (solid line) and run LX (dashed line). For clarity, we show the instantaneous $\alpha$ and the time-averaged $\alpha$ over a window of 20 local orbits, $\langle \alpha\rangle_{t,20}$. Run HX was initialized from the saturated state of LX, which explains the absence of an initial transient in HX.}%
    \end{figure}

The temporal evolution of the angular momentum transport coefficient $\alpha$ and turbulent poloidal energy $e_k$ is first considered; these are defined as
\begin{align}
	\alpha&=\overline{ \rho v_r \delta v_\phi}/\overline{P},\\
	e_k&=\frac{1}{2}\overline{\rho (v_r^2+v_\theta^2)}/\overline{P}.
\end{align}
These quantities, measured at $R=4$ in simulations LX, HX, and EX, are shown in Fig.~ \ref{fig:transport-t}. The LX simulation exhibits transient growth from 0 to saturation for the first $300$ orbits. This initial transient corresponds to the linear growth and saturation of the VSI. It is absent in the HX simulation because this run is initialised from the saturated state of LX. All simulations converge to the same steady state with $\langle \alpha\rangle_t=1.3\times 10^{-3}$ and $\langle e_K\rangle_t=1.9\times 10^{-2}$ when averaged over the $1000$ orbits, indicating that the bulk transport properties converge with resolution in these simulations. The instantaneous $\alpha$ oscillates widely between $\pm10^{-2}$, and only after applying a $20$-orbit running mean is the long-time average recovered.

\begin{figure}
   \centering
   \hspace{-5.0mm}\includegraphics[width=1.05\linewidth]{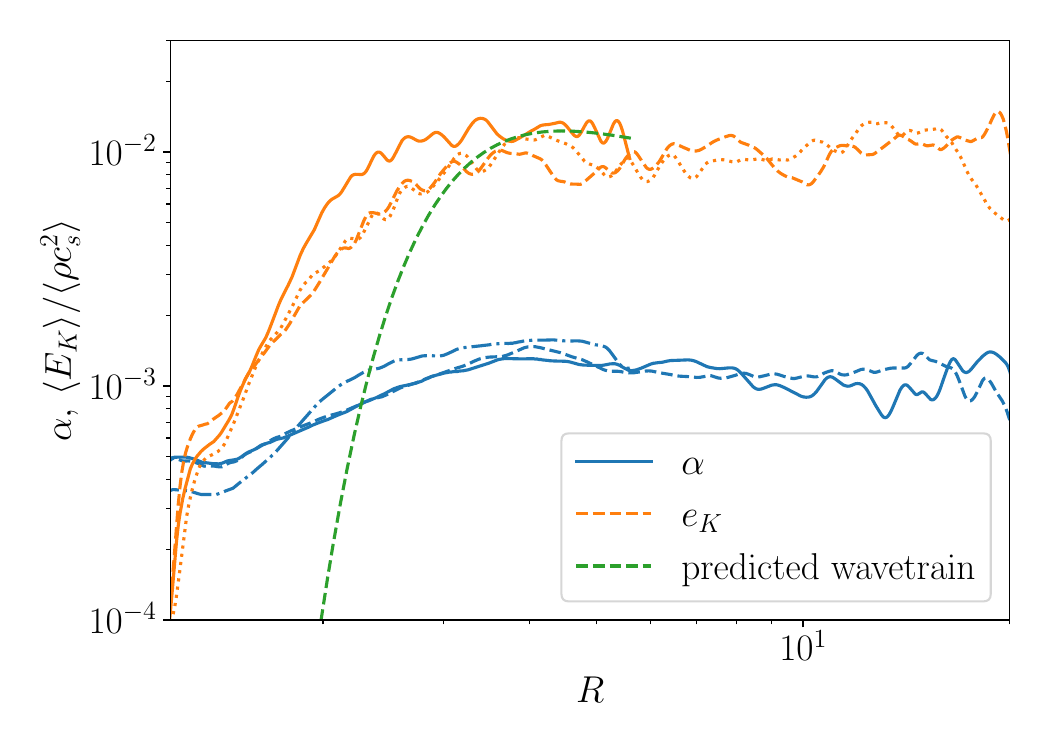}

   \caption{\label{fig:transport-R}Angular momentum transport and poloidal kinetic energy ($e_k$) averaged over the last 400 orbits of the simulations EX (solid line), HX (dashed line), and LX (dotted line). The predicted amplification and saturation radius for a single linear wave train with frequency $\omega=0.075\Omega_0$ and $n=1$ from \cite{Ogilvie.Latter.ea25} is shown as a dashed green line (eq.~\ref{eq:lin_wave train}).}%
    \end{figure}

Fig.~\ref{fig:transport-R} shows the radial dependence of the transport properties. We excised the radial buffer zones used to damp wave reflections to ensure that this figure accurately reflects the `active domain' of the simulations. The transport properties increase gradually between $R=1.2$ and $R=4$ where they reach a plateau. This feature is present in all of the models with the same amplitude, indicating convergence with spatial resolution. However, this behaviour is incompatible with the self-similar disc assumption, for which constant dimensionless values would be expected. This observation therefore likely indicates the influence of the inner boundary condition, which propagates deep ($\Delta R\sim 20H$) inside the active domain. For $R>4$, all models appear perfectly self-similar and show values similar to the time averages in Fig.~\ref{fig:transport-t}.

The average value of $\langle \alpha\rangle_t$ at $R=4$ (fig.~\ref{fig:transport-t}) is about twice the value reported in the 3D models of \cite{Shariff.Umurhan24} and approximately 30\% larger than that reported by \cite{Manger.Klahr18}. This discrepancy is probably due to the more extended radial domain used in our simulations. At $R=1.5$, $\langle \alpha\rangle_t\sim 5\times 10^{-4}$ , a value compatible with that reported by \cite{Manger.Klahr18}, but still significantly larger than \cite{Shariff.Umurhan24}.

\subsection{Flow structure\label{sec:flowStructure}}
\begin{figure}
   \centering

   \includegraphics[width=1.05\linewidth]{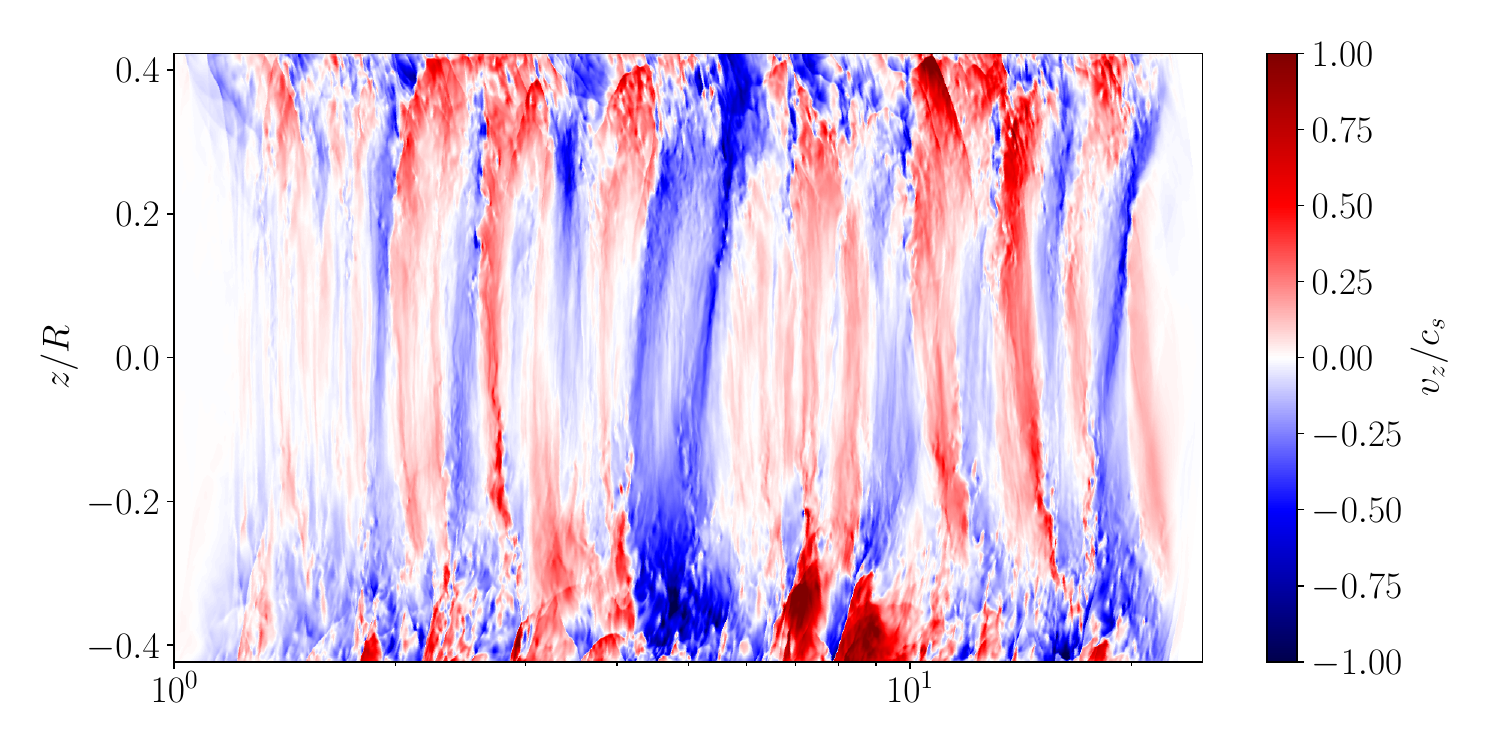}
   \includegraphics[width=1.05\linewidth]{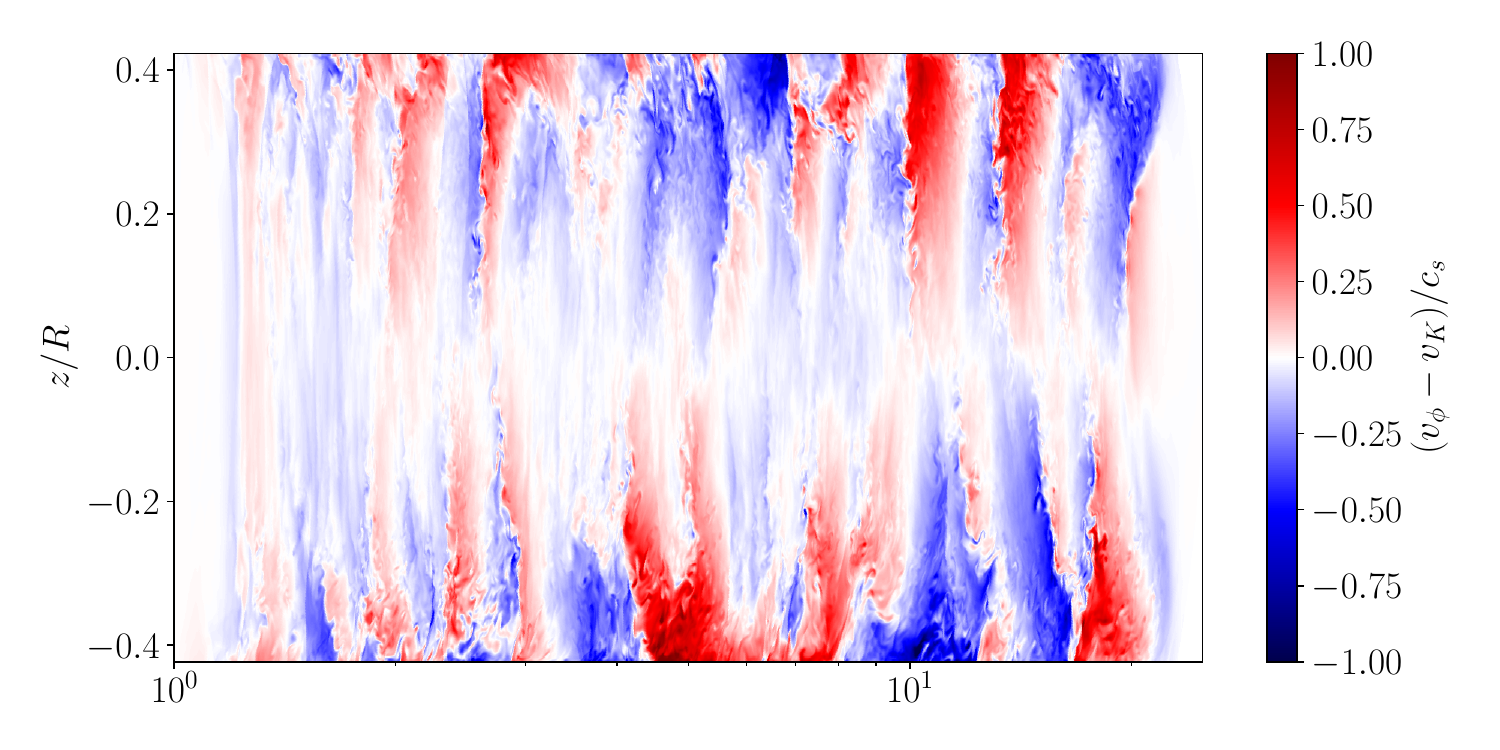}
   \includegraphics[width=1.05\linewidth]{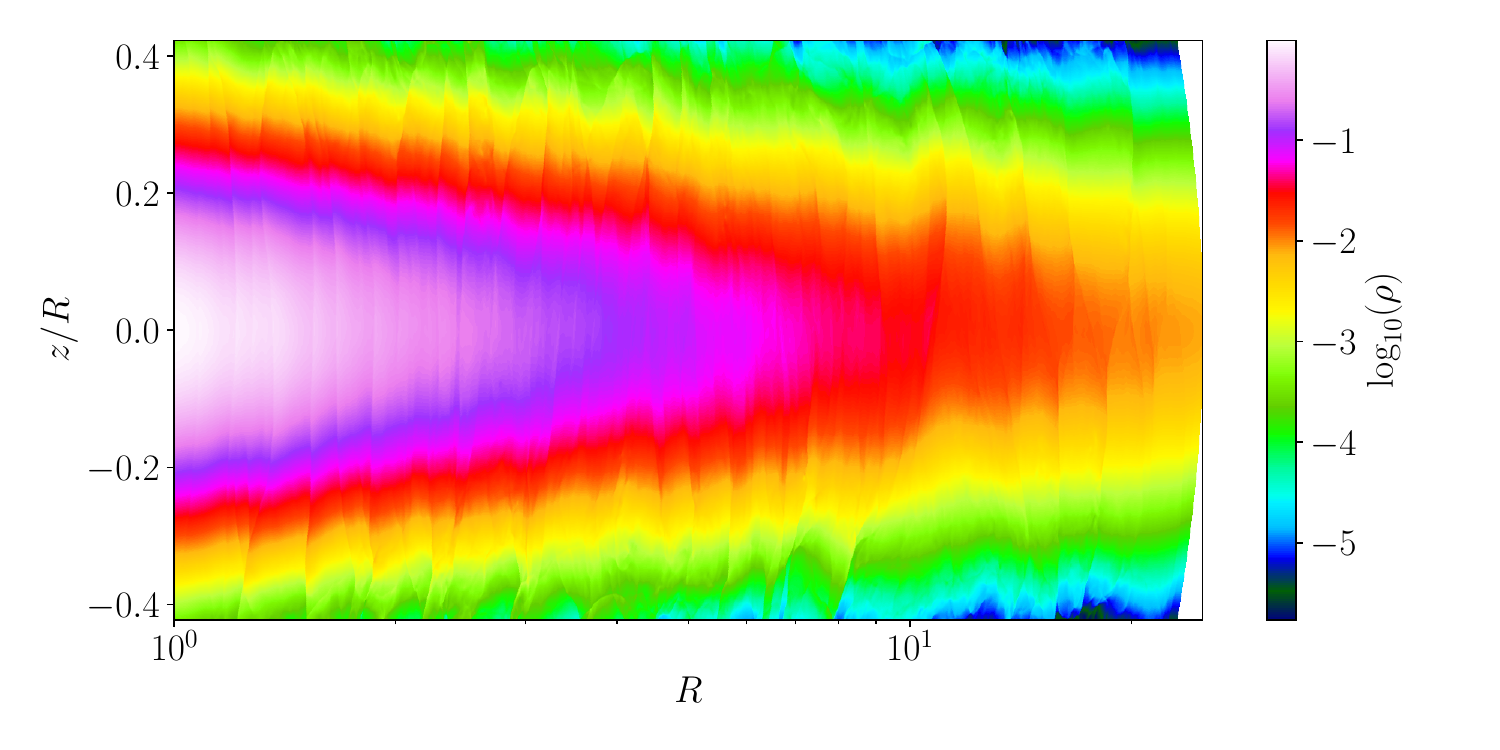}

   \caption{\label{fig:snapRz}Instantaneous $(R,z)$ cut of the HX model at $t=900$ orbits. Large-scale $n=1$ corrugation modes are evident in $v_z$, reaching sonic velocities (top panel), along with shock fronts propagating radially in the density field (bottom panel). }%
    \end{figure}

The instantaneous flow shape appears in an $R-z$ cut at a fixed azimuth (Fig.~\ref{fig:snapRz}). Large corrugation $n=1$ modes \citep{Ogilvie.Latter.ea25} are recovered, characterised by a column moving vertically up or down uniformly, alongside horizontal oscillations antisymmetric with respect to the disc midplane. In addition to these structures, we observe fine-scale structures and shocks, the latter being particularly evident in $v_R$ and $\rho$. The velocity magnitude is clearly sonic at three scale heights, which explains the presence of moderately strong ($\delta\rho/\rho\sim 1$) density jumps in the flow, especially visible in the disc atmosphere ($|z|\gtrsim 0.2R=2H$).

\begin{figure*}
   \centering
   \includegraphics[width=1.0\linewidth]{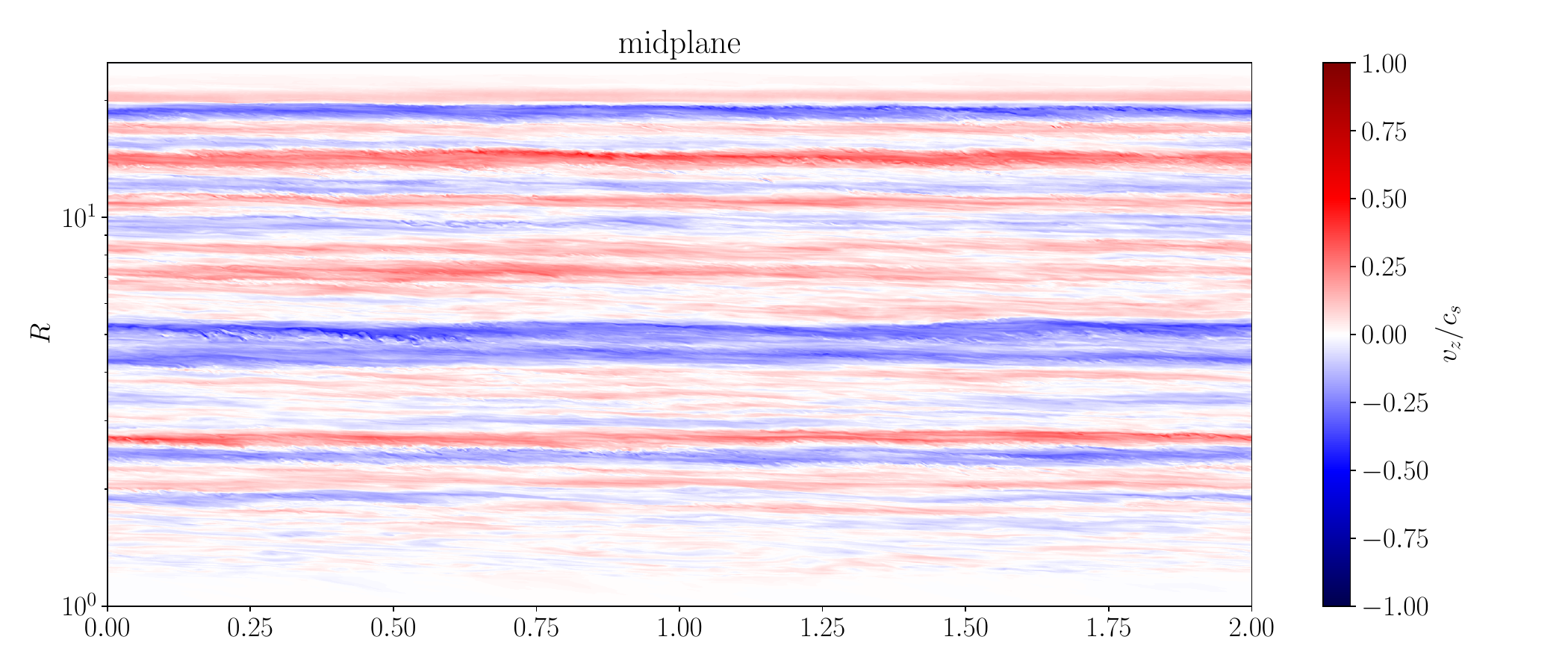}
   \includegraphics[width=1.0\linewidth]{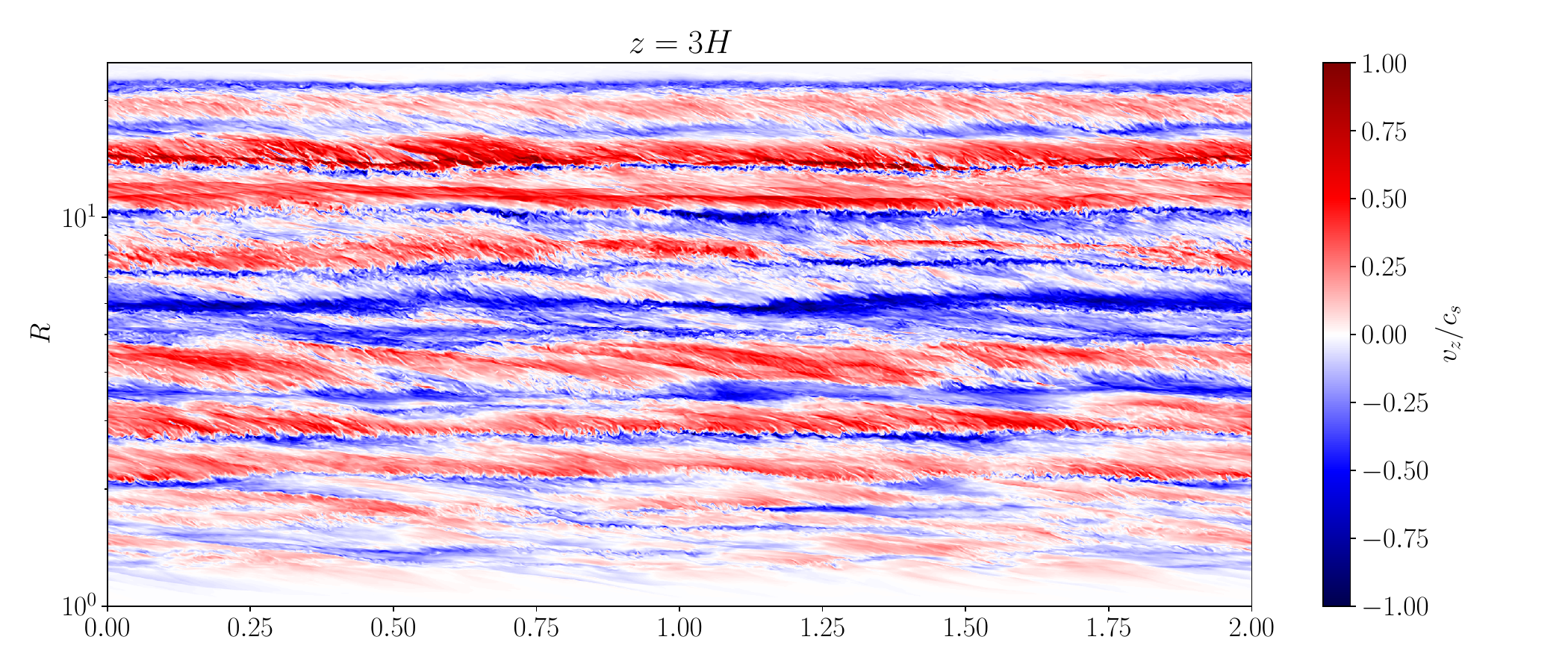}
   \includegraphics[width=1.0\linewidth]{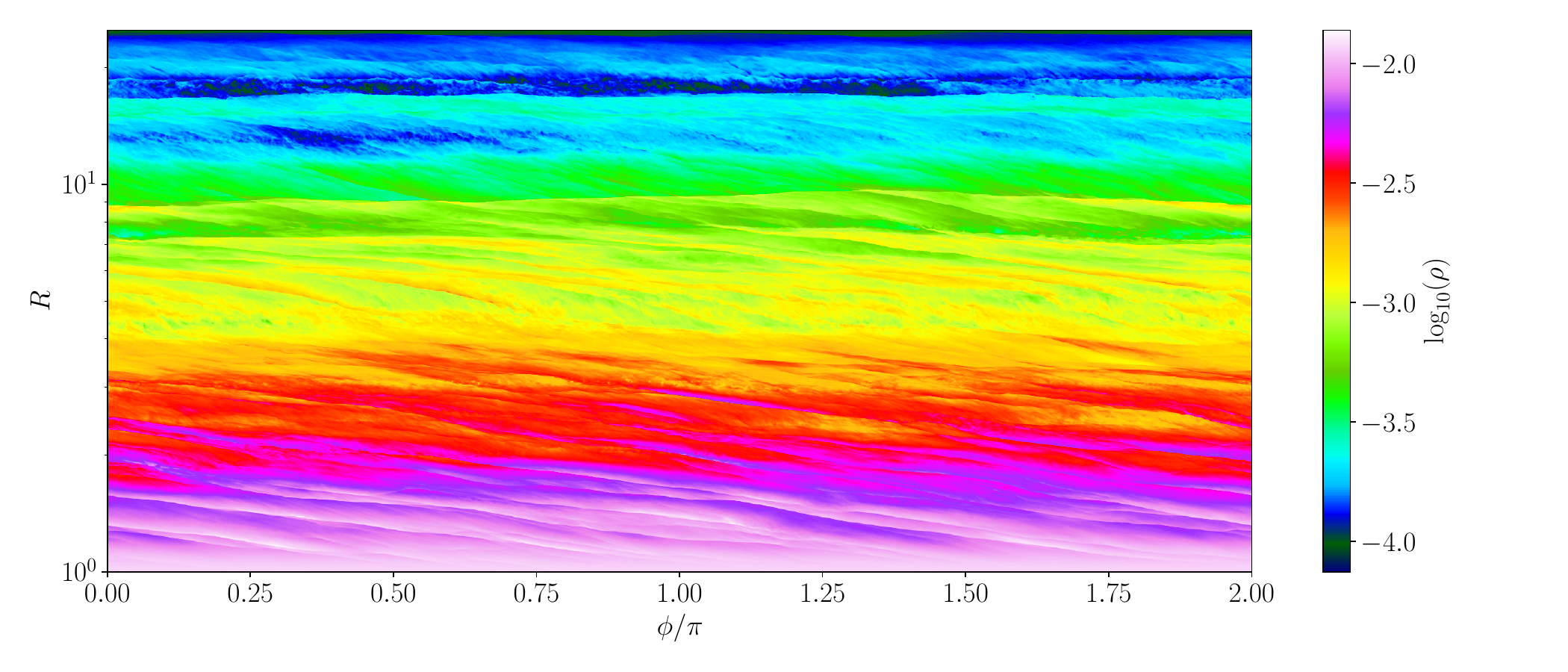}

   \caption{\label{fig:snapRphi}Instantaneous $(R,\phi)$ cuts of vertical velocity ($v_z$) and density ($\rho$) in the HX model at $t=900$ orbits. Top: Midplane cut of 
  $v_z$. Middle: $z=0.3R$ (corresponding to 3 pressure scale heights) cut of $v_z$. Bottom: $z=0.3R$ cut of the gas density $\rho$. }%
    \end{figure*}

The $R-\phi$ cuts (Fig.~\ref{fig:snapRphi}) reveal large-scale $v_z$ motions that remain remarkably axisymmetric in the disc midplane. In the disc atmosphere, at $3H$, large-scale vertical motion becomes stronger, but non-axisymmetric structures are also much more pronounced. Small-scale eddies, which resemble radial mixing of $v_z$, and oblique shocks, are distinguishable. These shocks become apparent in the density at $z=3H$ (Fig.~\ref{fig:snapRphi} bottom), where trailing non-axisymmetric shocks appear throughout the simulation domain.

\subsection{Zonal flows}

The appearance of long-lived zonal flows in the context of VSI turbulence has been proposed in several studies \citep[e.g.][]{Richard.Nelson.ea16,Manger.Klahr18}. Such zonal flows could prevent the inward migration of centimetre-sized dust grains, a problem historically known as the metre barrier (\citealt{Weidenschilling77}). For these zonal flows to act as efficient dust traps, they must generate pressure bumps; that is, the flow should become locally super-Keplerian.

\begin{figure}
   \centering
\includegraphics[width=1.1\linewidth]{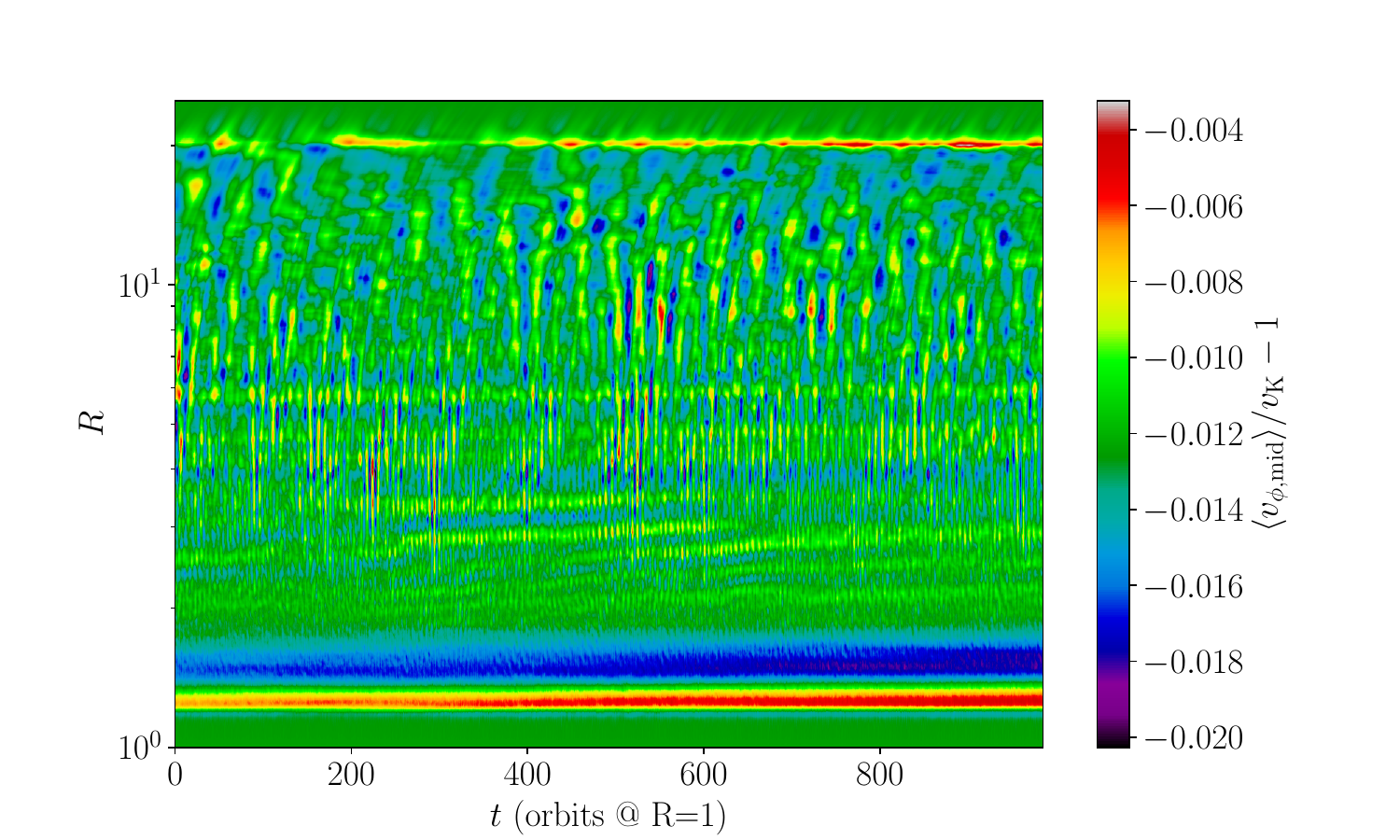}
   \caption{\label{fig:spacetime_vphi}Spatio-temporal evolution of the azimuthally averaged midplane $\delta v_\phi /v_K$ in run HX. Weak zonal flows are observed, with deviations from Kepler of the order of 1\%. The flow remains sub-Keplerian in the midplane throughout the run.}%
    \end{figure}

Fig.~\ref{fig:spacetime_vphi} presents the relative deviation from the Keplerian velocity in the disc midplane for run HX, where two strong zonal flows appear close to the inner ($R\simeq 1.2$) and outer ($R\simeq 21$) radial boundaries. These features are clearly boundary condition artefacts and should be disregarded. In the bulk of the disc, weak deviations from Keplerian rotation are present but never exceed $1\%$, and the flow remains on average Keplerian.

This absence of strong zonal flows may be related to the flow symmetry with respect to the disc midplane (Fig.~\ref{fig:snapRz}). In particular, we find that $\delta v_\phi$ is always antisymmetric with respect to $z=0$. This antisymmetry of $\delta v_\phi$, together with the symmetry of $v_z$, is characteristic of the $n=1$ corrugation mode \citep{Ogilvie.Latter.ea25}, and we confirm this mode selection in section \ref{sec:corrugation}. These observations indicate that deviations from Keplerian rotation in the disc midplane are weak and insufficient to feed long-lived pressure bumps or dust traps.

\subsection{\label{sec:corrugation}Large-scale wave propagation}

\begin{figure}
   \centering
\includegraphics[width=1.1\linewidth]{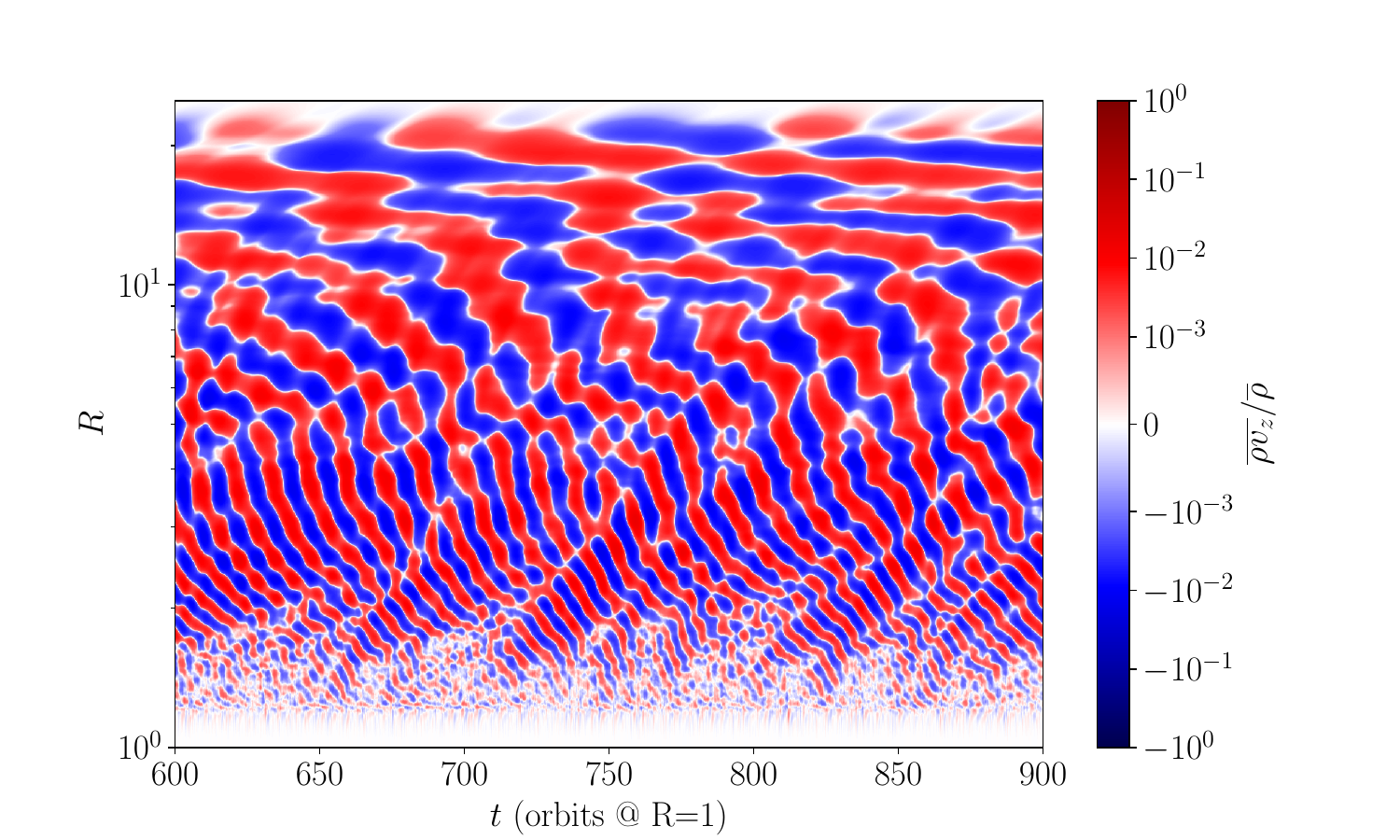}
   \caption{\label{fig:spacetime_vtheta}Spatio-temporal evolution of the mean vertical mass flux averaged over $\phi$ and $\theta$ in simulation HX. Characteristic wave patterns of large-scale corrugation modes, first identified in 2D simulations, are recovered.}%
    \end{figure}

As discussed above, we recover the large-scale $n=1$ corrugation modes identified in 2D simulations \citep{Stoll.Kley14, Svanberg.Cui.ea22}. These are clearly visible in the space-time diagram of the mean vertical mass flux $\overline{\rho v_z}$ (Fig.~\ref{fig:spacetime_vtheta}). The diagram reveals several wave zones with temporal and spatial frequencies. 

\begin{figure}
   \centering
\includegraphics[width=1.1\linewidth]{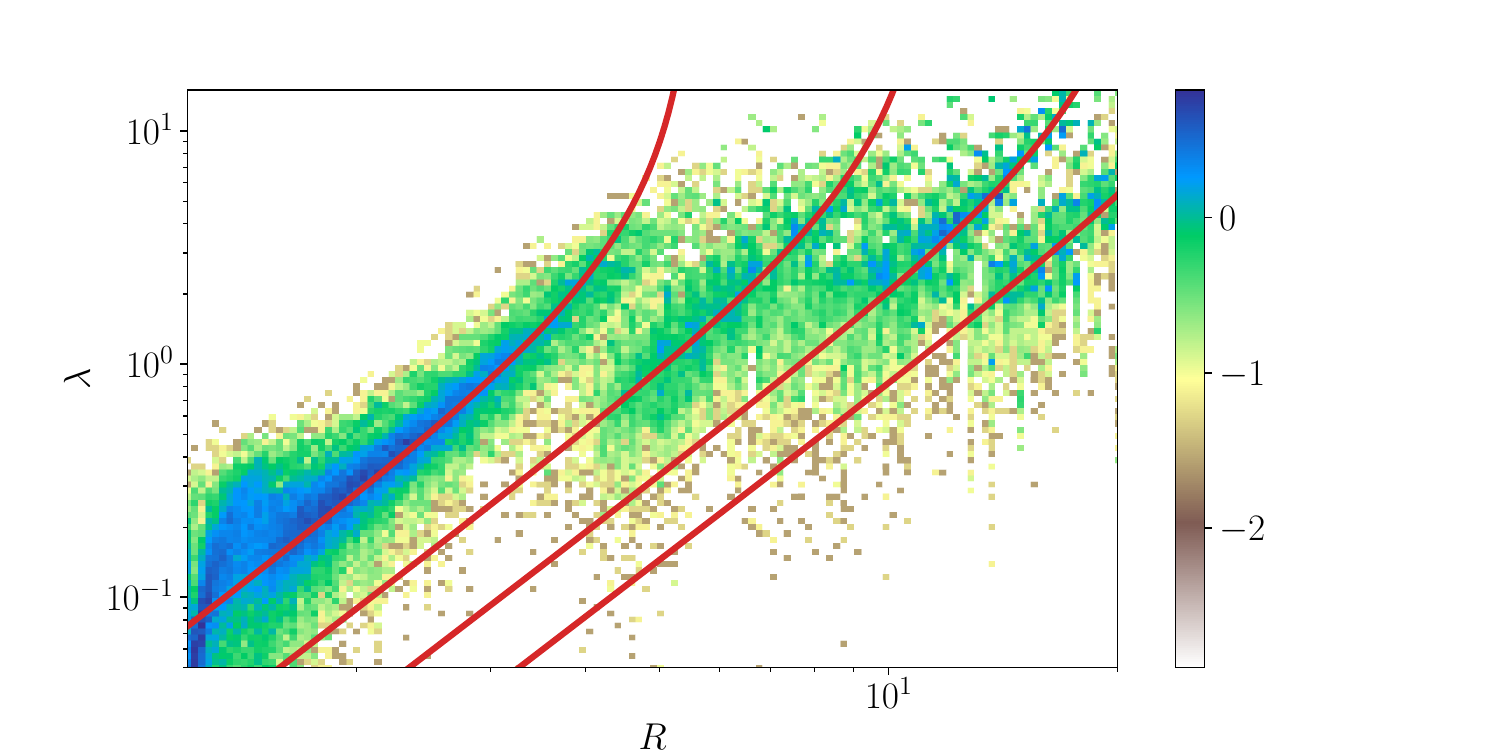}
   \caption{\label{fig:wavelength}Radial wavelength probability distribution of the large-scale vertical corrugation motion versus R in run HX, computed over the entire simulation duration. Over-plotted in red are the expected frequency of $n=1$ inertial modes for frequencies $\omega/\Omega_0=0.075,\,0.025,\,0.0094,\text{and}\,0.0041$. }%
    \end{figure}

To better diagnose these properties, we first measured the radial wavelength of the corrugation pattern between two nulls of $\overline{\rho v_z}(r,t)$ at each time in the HX simulation. This provided us with a distribution of radial wavelengths at each radius (Fig.~\ref{fig:wavelength}). The figure reveals four wave regions,  each following the linear dispersion relation of an $n=1$ inertial mode with a fixed frequency $\omega$:
\begin{align}
\label{eq:inertial_modes}
(\omega^2-n\Omega^2)(\omega^2-\Omega^2)=(\omega c_\text{s}k)^2,
\end{align}
where $\Omega$ is the local Keplerian frequency, $c_s$ is the local isothermal sound speed, and $k=2\pi/\lambda$ is the radial wavenumber \citep{Lubow.Pringle93, Ogilvie.Latter.ea25}. We fit the wave regions in Fig.~\ref{fig:wavelength}, using the frequencies $\omega/\Omega_0=0.075,0.025,0.0094$, and $0.0041$ (where $\Omega_0\equiv\Omega(R=1)$), representing a best fit `by eye' to the data. These inertial waves appear when $\omega<\Omega$ and possess a Lindblad resonance at $R=R_L$, which satisfies $\omega=\Omega(R_L)$; beyond this point, they become (stable) acoustic waves. In our simulations, the wave trains tend to vanish as they approach their respective resonances and are replaced by lower-frequency wave trains. A very steep band of short waves near the inner boundary provides further evidence of inner boundary artefacts.

\begin{figure}
   \centering
\includegraphics[width=1.1\linewidth]{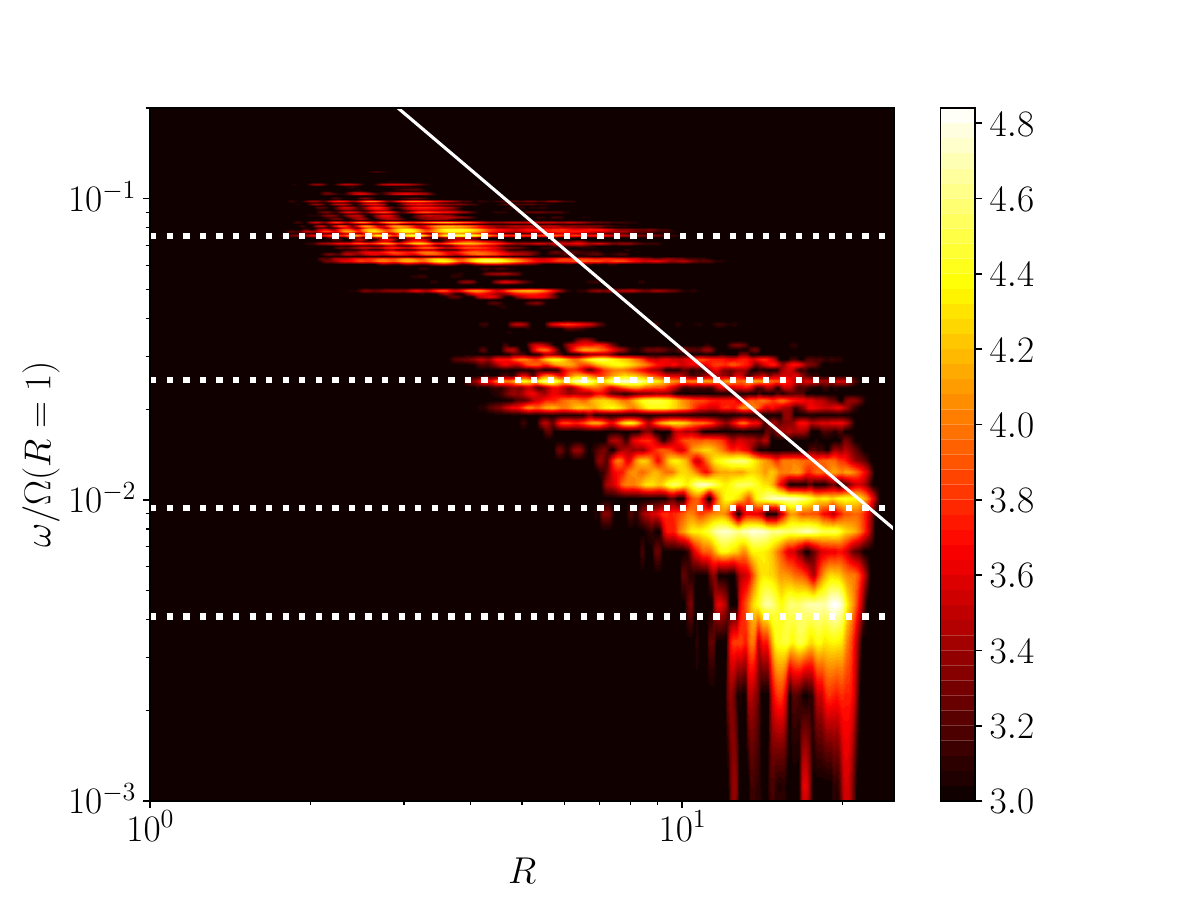}
   \caption{\label{fig:temporal-spectrum}Temporal spectrum of the large-scale vertical corrugation motion versus $R$ in run HX. The dotted lines show the frequencies of the inertial modes plotted in Fig.~\ref{fig:wavelength} ($\omega/\Omega_0=0.075,\,0.025,\,0.0094,\text{and}\,0.0041$), and the solid line indicates the Keplerian frequency $\omega=\Omega_K$. }%
    \end{figure}

 Fig.~\ref{fig:temporal-spectrum}. shows the directly measured temporal spectrum of the corrugation waves. The spectrum displays horizontal bands grouped into broader structures. Each horizontal band corresponds to a particular wave frequency excited over a range of radii. The dotted line corresponds to the best-fit frequencies mentioned above and approximately fits the large bumps in the structures. However, the temporal spectrum is much broader in frequency than in 2D \citep{Svanberg.Cui.ea22}. This is possibly due to small-scale turbulence,  which excites additional modes of flow. Surprisingly, we recover the same wave frequencies (within 10\%) as those found in 2D simulations, indicating that the VSI somehow selects those particular frequencies.

It is tempting to associate the increase in kinetic energy and transport (Fig.~\ref{fig:transport-R}) observed over $1<R<4$ with the amplification of the first wave train. Using a linear analysis, \cite{Ogilvie.Latter.ea25} proposed that the amplitude of a single wave train increases with $R$ according to
\begin{align}
\label{eq:lin_wave train}
e_k\propto \left(1+\left(\frac{R_L}{R}\right)^3\right)\exp\left(-\frac{R_L^3}{3R^3}\right).
\end{align}
Fig.~\ref{fig:transport-R} shows this prediction for the frequency of our first wave train and reveals that linear theory largely overestimates the radial growth of the wave train. This mismatch may originate from (1) the wave train having already saturated non-linearly and thus is subject to a non-linear viscous damping due to non-axisymmetric structures (cf.~section \ref{sec:turbulent-viscosity}), which is absent from the linear analysis of \cite{Ogilvie.Latter.ea25}; or (2) the inner wave train is not strictly monochromatic (see Fig.~\ref{fig:temporal-spectrum}), and we are witnessing a combination of radial growth from a spread of wave trains. The radius at which $e_k$ saturates matches the plateau of the linear theory. Whether this match constitutes a coincidence or a success of the linear theory remains unclear; a more systematic exploration of the parameter space is necessary to confirm this conjecture.

\subsection{Large-scale vortices}
   
   As discussed in \ref{sec:flowStructure}, the large-scale corrugation motions are accompanied by small-scale structures, which are particularly visible at $3H$ (Fig.~\ref{fig:snapRphi} middle), but are also present in the disc midplane. A key question for planet formation theories is the nature of these small-scale structures and the potential emergence of large-scale vortices, which may aid in the assembly of planetary embryos. Such vortices have been found in the 3D simulations of \cite{Richard.Nelson.ea16} and were attributed to a local manifestation of the Rossby wave instability \citep[RWI;][]{Lovelace.Li.ea99}. 
   
\begin{figure}
   \centering
\includegraphics[width=1.1\linewidth]{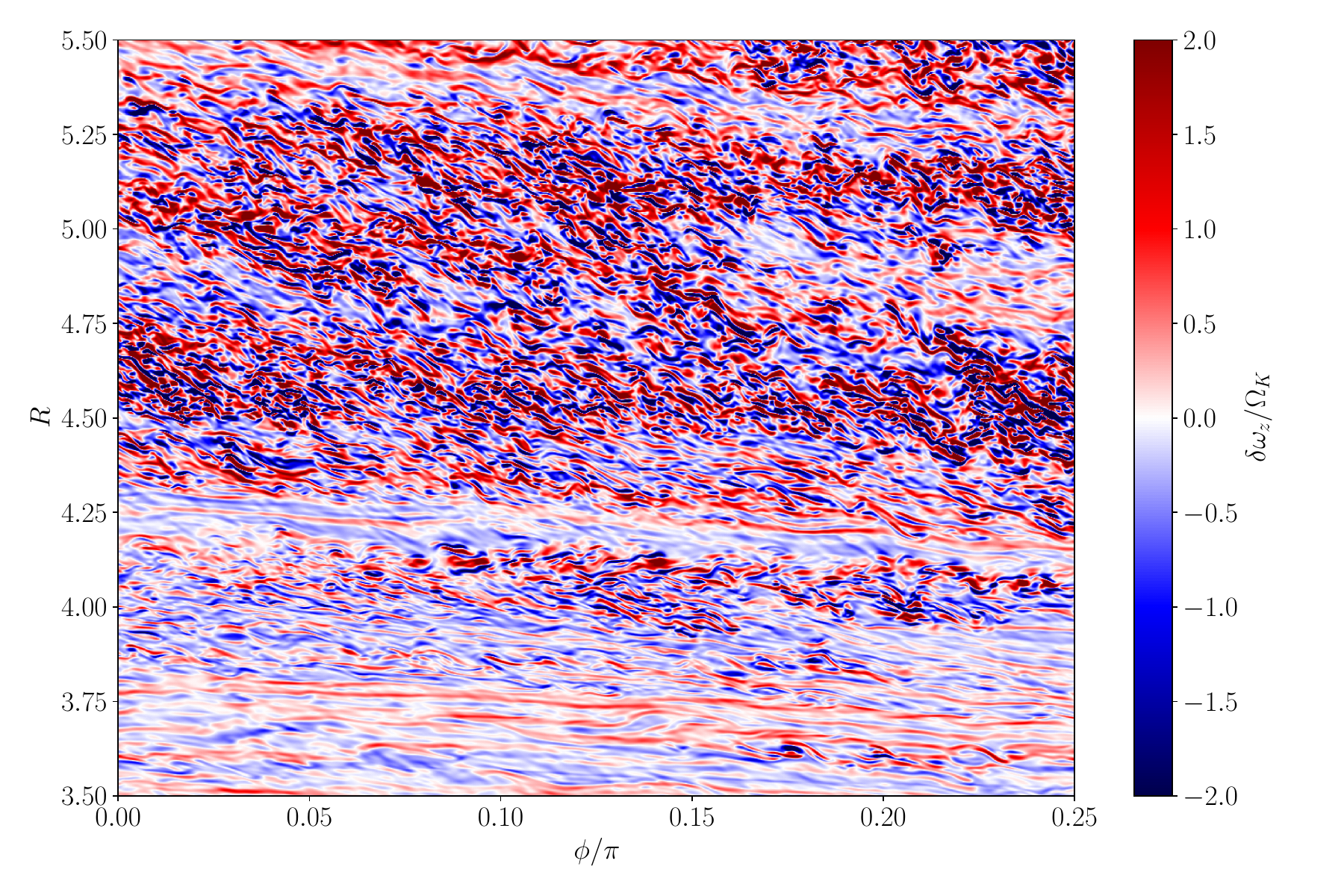}
   \caption{\label{fig:wz-midplane}Snapshot of the vertical vorticity perturbation in simulation EX at $t=800$ orbits in the disc midplane, zooming in of the region around $R=4.5$. The image illustrates the complex turbulent flow and the absence of any large-scale coherent vortical structures.}%
    \end{figure}

  We therefore searched for large-scale vortices in the disc midplane. The midplane appears remarkably quiet in terms of vertical velocity (Fig.~\ref{fig:snapRphi}, top), with most chaotic motions occurring in the disc atmosphere. A standard tracer for vortices is the deviation in vertical vorticity $\delta \omega_z=r^{-1}[ \partial_r(r\,\delta v_\phi)-\partial_\phi \delta v_r]$, where the background vorticity of the mean Keplerian shear was subtracted. The vertical vorticity is shown in Fig.~\ref{fig:wz-midplane} and reveals numerous small-scale fluctuations that are much smaller than the disc scale height. These fluctuations are highly dynamical (i.e. they evolve on timescales shorter than the local orbital timescale) and do not exhibit any preferential direction with respect to the disc rotation axis. These results stand in sharp contrast to those of \cite{Richard.Nelson.ea16} and \cite{Manger.Klahr18} (hereafter MK18), who both observed the formation of long-lived anticyclonic vortices (i.e., $\delta \omega_z\Omega<0$) at scales $\sim H$. 
  
  The reason for the absence of long-lived vortices in the high-resolution models is not entirely clear, but several hypotheses can be proposed. First, the resolution is much higher than in \cite{Richard.Nelson.ea16} (50 points/$H$, depending on the direction) and in MK18 (20 points/$H$). In addition, we used a high-order reconstruction scheme (LimO3) and orbital advection with parabolic reconstruction to minimise numerical dissipation. As a result, our models are able to capture thinner structures that could disrupt large-scale vortices, such as elliptical instabilities \citep{Lesur.Papaloizou09}. A similar absence of vortices is observed in the high-resolution models (80 points/$H$) of \cite{Shariff.Umurhan24}, whereas \cite{Huang.Bai25} report vortices at lower resolution (30 points/$H$), supporting the idea of a resolution effect. A second point of difference lies in the cooling timescale, which is infinitely small, as we employed the locally isothermal approximation. \citet{Richard.Nelson.ea16} observed that the vortex aspect ratio and lifetime both increase with increasing cooling time. We might therefore be witnessing the breakup of the strongest vortices found by \cite{Richard.Nelson.ea16}, while weaker vortices at longer cooling timescales (not explored here) might still be present at high resolution. 
  
\subsection{Spectra of the small-scale turbulence}

\begin{figure}
   \centering
\includegraphics[width=1.0\linewidth]{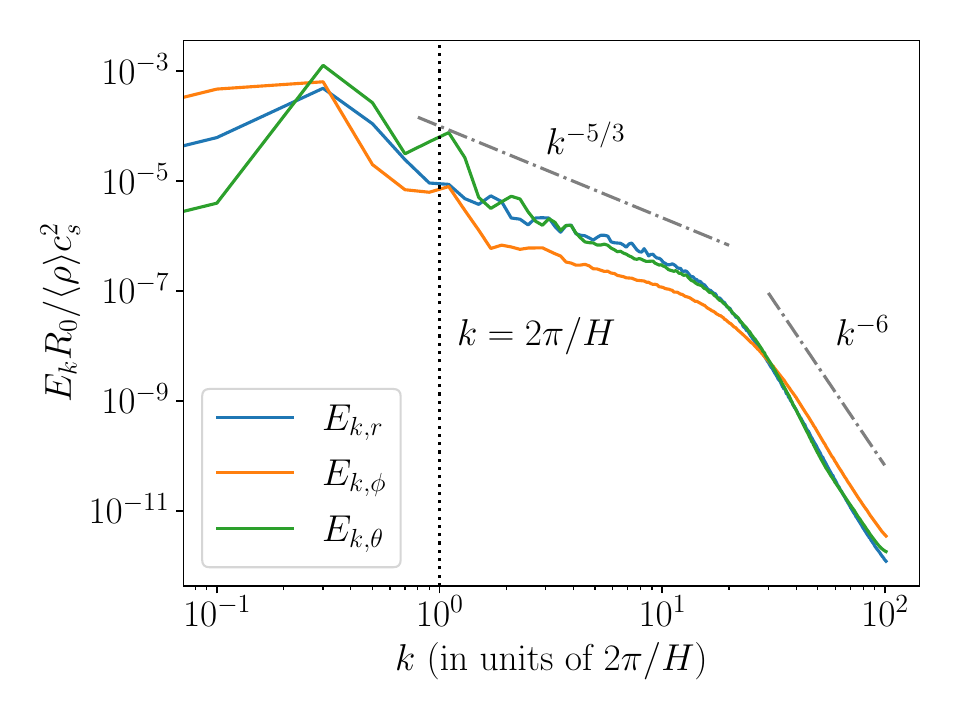}
   \caption{\label{fig:1D-spectrum}Power spectrum of kinetic energy obtained by integrating the 3D spectrum over spherical shells for each velocity field component at the middle of the computational domain in run EX (see text) . }%
    \end{figure}

  The fine-scale structures, being 3D and transient, can be interpreted as manifestations of small-scale ($<H$) turbulence. To explore this idea, we assessed whether this turbulence can be described within a `standard' fluid-turbulence framework. To this end, we computed the 3D shell-integrated spectrum of run EX, shown in Fig.~\ref{fig:1D-spectrum}. This spectrum was obtained by extracting a cubic box from run EX centred on $R_0=4$ spanning $r\in [3.0,5.0]\,;\,\theta-\pi/2\in [-0.2,+0.2]   \,;\,\phi\in[0,0.6]$, running a 3D Fast Fourier transform using a Hanning window, and computing the total kinetic energy in each spherical shell of the resulting spectrum. This procedure allowed us to capture the spectrum over three decades in wavelength, at locations sufficiently far from the radial and vertical boundaries, within a box approximately $\pm 2H$ in size centred around $R_0=4$ on the disc midplane. The turbulence spectrum essentially follows a $k^{-5/3}$ power law between $k=1$ (i.e.,\ the disc scale height) and $k\sim 15$, beyond which the spectrum steepens, with an approximate $k^{-6}$ dependence. Additionally, a strong peak is observed in $E_\theta$ at the local scale height wavenumber $k\sim 1$, which is a signature of the large-scale corrugation modes, acting as the injection scale for the system.    

\begin{figure}
   \centering
\includegraphics[width=1.0\linewidth]{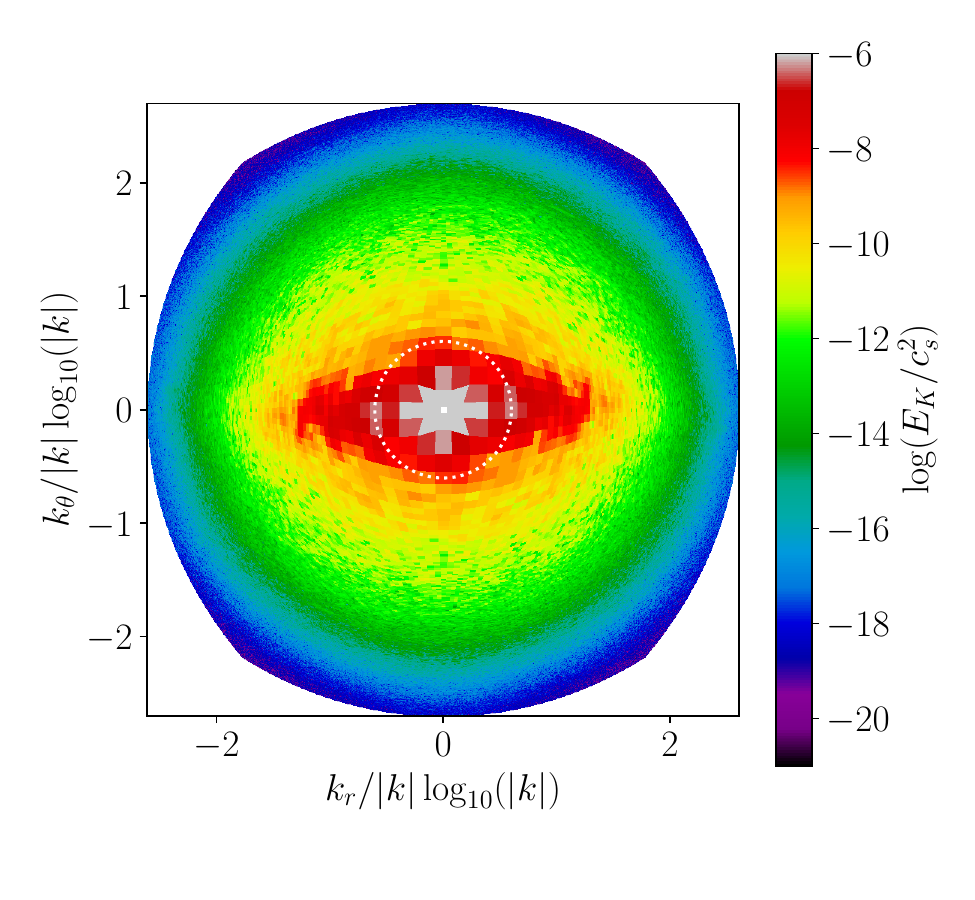}
   \caption{\label{fig:2D-spectrum-r-theta}Two-dimensional cut in the $(k_r,k_\theta)$ plane at $k_\phi=0$ through the 3D spectrum of the total kinetic energy in run EX. The dotted white line denotes the disc scale height.}%
    \end{figure}

\begin{figure}
   \centering
\includegraphics[width=1.0\linewidth]{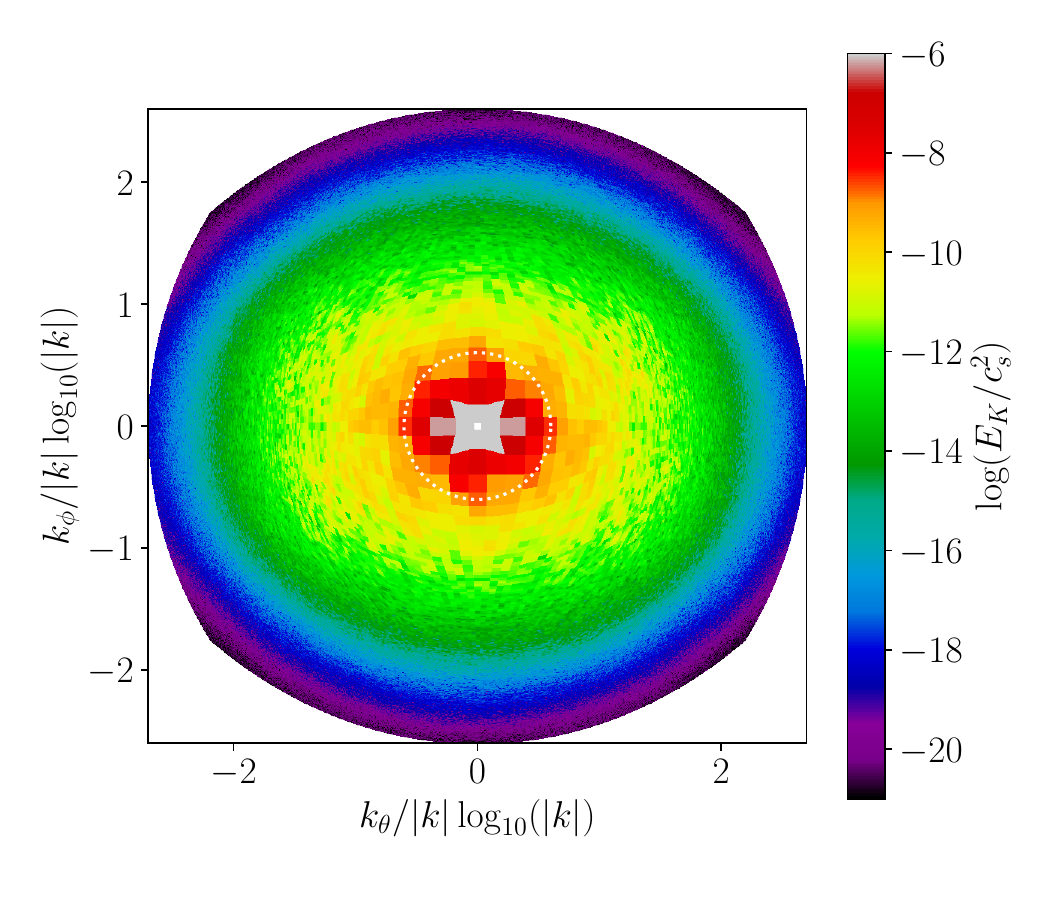}
   \caption{\label{fig:2D-spectrum-theta-phi}Two-dimensional cut in the $(k_\theta,k_\phi)$ plane at $k_r=0$ through the 3D spectrum of the total kinetic energy in run EX. The dotted white line denotes the disc scale height. This cut is representative of the isotropy observed for $k_r\ne 0$ and is not a peculiarity of the $k_r=0$ plane. }%
    \end{figure}

   Beyond 1D shell-integrated spectra, we examined turbulence isotropy. We characterised this property using 2D spectra, i.e., slices through the 3D spectrum of the domain. Figure\ref{fig:2D-spectrum-r-theta} shows a cut through the $(k_r,k_\theta)$ plane at $k_\phi=0$. This spectrum can therefore be interpreted as an `axisymmetric' spectrum. A striking feature is the presence of two `hot spots' localised at $k_\theta\simeq 0$ and $k_r\sim 2\pi/H$ (the scale height being identified as a dotted white line). These are a clear signature of the large-scale $n=1$ corrugation wave, with this hot spot being dominated by energy in vertical motions (see Fig.~\ref{fig:2D-spectrum-full} in the Appendix). In addition, we observe a long tail of energy in the $k_r$ direction, which makes the spectrum strongly anisotropic. This tail is associated either with the non-linear radial steepening of the corrugation wave, which essentially creates smaller radial scales, or simply with rotation, which is known to produce anisotropic turbulence with structures elongated along the rotation axis. This steepening eventually feeds the cascade in the $\theta$ and $\phi$ directions, likely through secondary shear instability. Fig.~\ref{fig:2D-spectrum-theta-phi} clearly shows this effect in the $(k_\theta,k_\phi)$ plane, where the spectrum is now essentially isotropic.

\begin{figure}
   \centering
\includegraphics[width=1.0\linewidth]{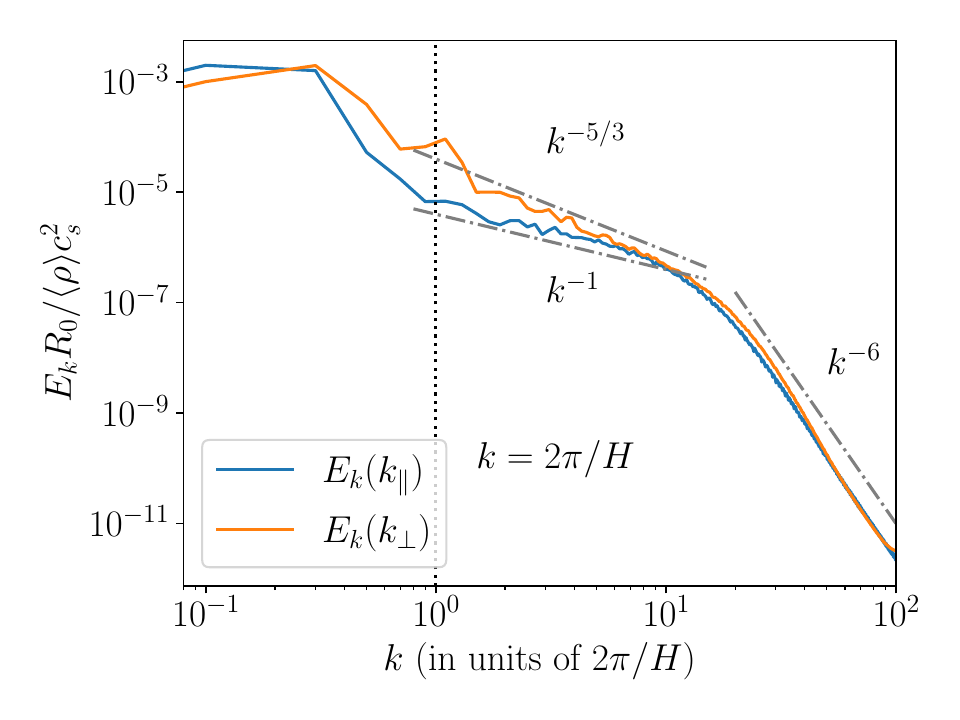}
   \caption{\label{fig:1D-spectrum-perp}Power spectrum of kinetic energy along $k_\parallel$ and $k_\perp$ and power laws expected for critically balanced rotating turbulence, assuming $k_\perp=k_r$ and $k_\parallel=(k_\theta^2+k_\phi^2)^{1/2}$ (see text).  }%
    \end{figure}

The interpretation of the turbulent energy spectrum in our simulations can be approached through several theoretical frameworks: the classical Kolmogorov cascade \citep{Kolmogorov41}, wave turbulence theory \citep[e.g.][]{Nazarenko11}, and the framework of critically balanced turbulence, originally developed for magnetohydrodynamic (MHD) turbulence \citep{Goldreich.Sridhar95} and later extended to rotating flows \citep{Nazarenko.Schekochihin11}. The Kolmogorov cascade assumes homogeneous and isotropic turbulence and predicts an energy spectrum scaling as $E(k)\propto k^{-5/3}$. Although the spectral slope observed in our data (Fig.~\ref{fig:1D-spectrum}) is broadly consistent with this prediction, the pronounced anisotropy in the flow (Fig.~\ref{fig:2D-spectrum-r-theta}) clearly violates the Kolmogorov assumptions. In contrast, the spectrum expected for wave turbulence follows a steeper scaling of $E(k)\propto k^{-5/2}$, which is incompatible with our results and can therefore be confidently excluded. This leaves critical balance as the most appropriate interpretive framework. In this scenario, the turbulent cascade is governed by a balance between the linear wave propagation timescale and the nonlinear interaction timescale. This regime naturally gives rise to spectral anisotropy, characterised by different slopes along directions parallel and perpendicular to the rotation axis: $E(k_\parallel)\propto k_\parallel^{-1}$ and $E(k_\perp)\propto k_\perp^{-5/3}$, respectively, with non-linear interactions occurring primarily in the perpendicular direction \citep{Nazarenko.Schekochihin11, Nazarenko11}. To test this critical balance hypothesis in the context of the VSI, we computed 1D energy spectra along $k_\perp\equiv k_r$ and $k_\parallel\equiv(k_\theta^2+k_\phi^2)^{1/2}$ (Fig.~\ref{fig:1D-spectrum-perp}). The resulting anisotropic spectra support the interpretation of VSI turbulence as critically balanced rotating turbulence: the spectrum follows approximately $E(k_\perp)\propto k_r^{-5/3}$, while it is shallower with approximately $E(k_\parallel)\propto k_\parallel^{-1}$, until the two match at $k_i\sim 20$ and the spectrum becomes isotropic and much steeper. This result is not necessarily surprising, since the VSI is primarily an instability of inertial waves, while critically balanced rotating turbulence describes a cascade of strongly interacting inertial waves. It is, however, not an exact match, since the shear prevents the usual $k_\perp$ non-linear cascade from developing along $k_\phi$. Hence, in our case, $k_\perp$ lies only along $k_r$, while $k_\parallel$ also retains the shearwise $\phi$ direction, implying that the shearing of non-axisymmetric structures is a component of the linear wave propagation in the critical balance framework. Critical balance is expected to break down when the Rossby number $\delta \omega_z/\Omega\simeq 1$. We find that this is approximately the case around $k\sim 10$, as the vorticity `patches' in Fig.~\ref{fig:wz-midplane} have a size of approximately $H/10$ with $\delta \omega_z/\Omega=O(1)$.  We also note that despite the great care taken to limit numerical dissipation, we are still unable to see a proper isotropic Kolmogorov cascade, which would be expected for $k>20$, but instead find the much steeper $k^{-6}$ spectrum. It is unclear whether this steep spectrum is a genuine physical result or if we are witnessing the effect of numerical diffusion.

We note that a similar broken spectrum was obtained by MK18, with a break from $k^{-5/3}$ to $k^{-5}$ that they associate with a 2D inverse cascade. A similar phenomenon was observed in 2D high-resolution models by \cite{MelonFuksman.Flock.ea24a}, with a break to $k^{-7}$ that they propose could also be due to numerical diffusion. While it is tempting to think we are witnessing the same phenomenology as MK18, it is worth stressing that our break is at $k\simeq 15 $, while the break of MK18 is at $k\simeq 0.7$ (our choice of units for $k$ being equivalent to their $m=20\pi k=40$). Hence, our break occurs on a scale about 20 times smaller than that of MK18. Interestingly, MK18 used a resolution of $12$ points per $H$, i.e., about 20 times less than our EX simulation. This suggests that the break length scale follows the resolution and hence that the $k^{-6}$ part of the spectrum is probably due to numerical diffusion.

\subsection{Impact of small-scale turbulence on the large scales \label{sec:turbulent-viscosity}}

\begin{figure}
   \centering
\includegraphics[width=1.0\linewidth]{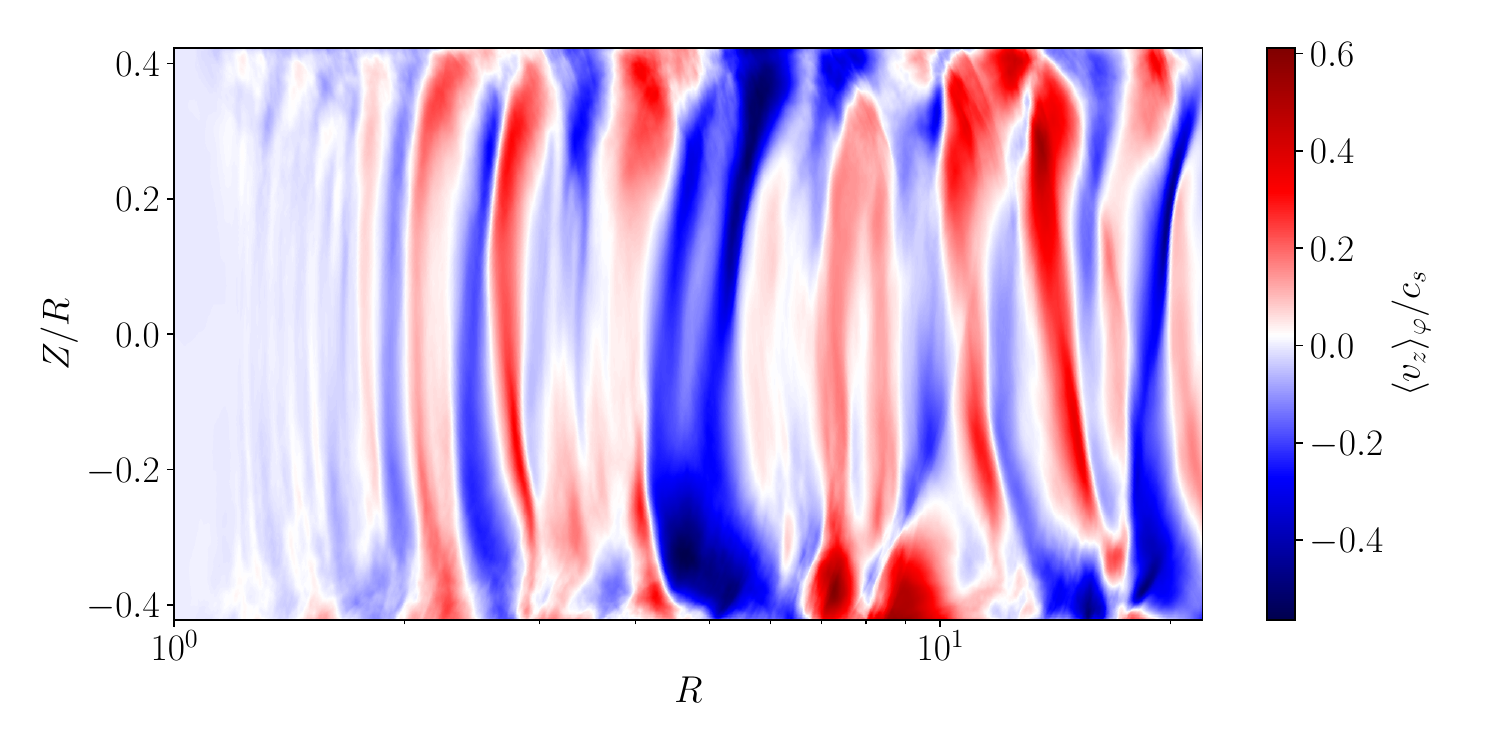}
   \caption{\label{fig:snap-vz-avg}Snapshot of $\phi$-averaged vertical velocity ($v_z$) in run HX at $t=900$ orbits.}%
    \end{figure}

A key question regarding the VSI concerns how the instability saturates. Previous studies propose that parametric instabilities could play a role \citep{Cui.Latter22}, while Rossby Wave instabilities \citep{Richard.Nelson.ea16} or Kelvin--Helmholtz instabilities \citep{Latter.Papaloizou18} could also contribute. An useful element in this discussion is the comparison of $v_z$ structures between the cut in the $(R,z)$ plane (Fig.~\ref{fig:snapRz}, second panel) and the $\phi-$averaged flow taken at the same time (Fig.~\ref{fig:snap-vz-avg}). The $\phi$-averaged flow recovers the large-scale corrugation modes but is much smoother than the $\phi=0$ cut; the small-scale fluctuations visible in Fig.~\ref{fig:snapRz} disappear, and therefore must have developed some degree of non-axisymmetry. This suggests that the $\phi-$averaged flow could be treated as a `viscous' 2D version of the full 3D model.

\begin{figure}
   \centering
\includegraphics[width=1.0\linewidth]{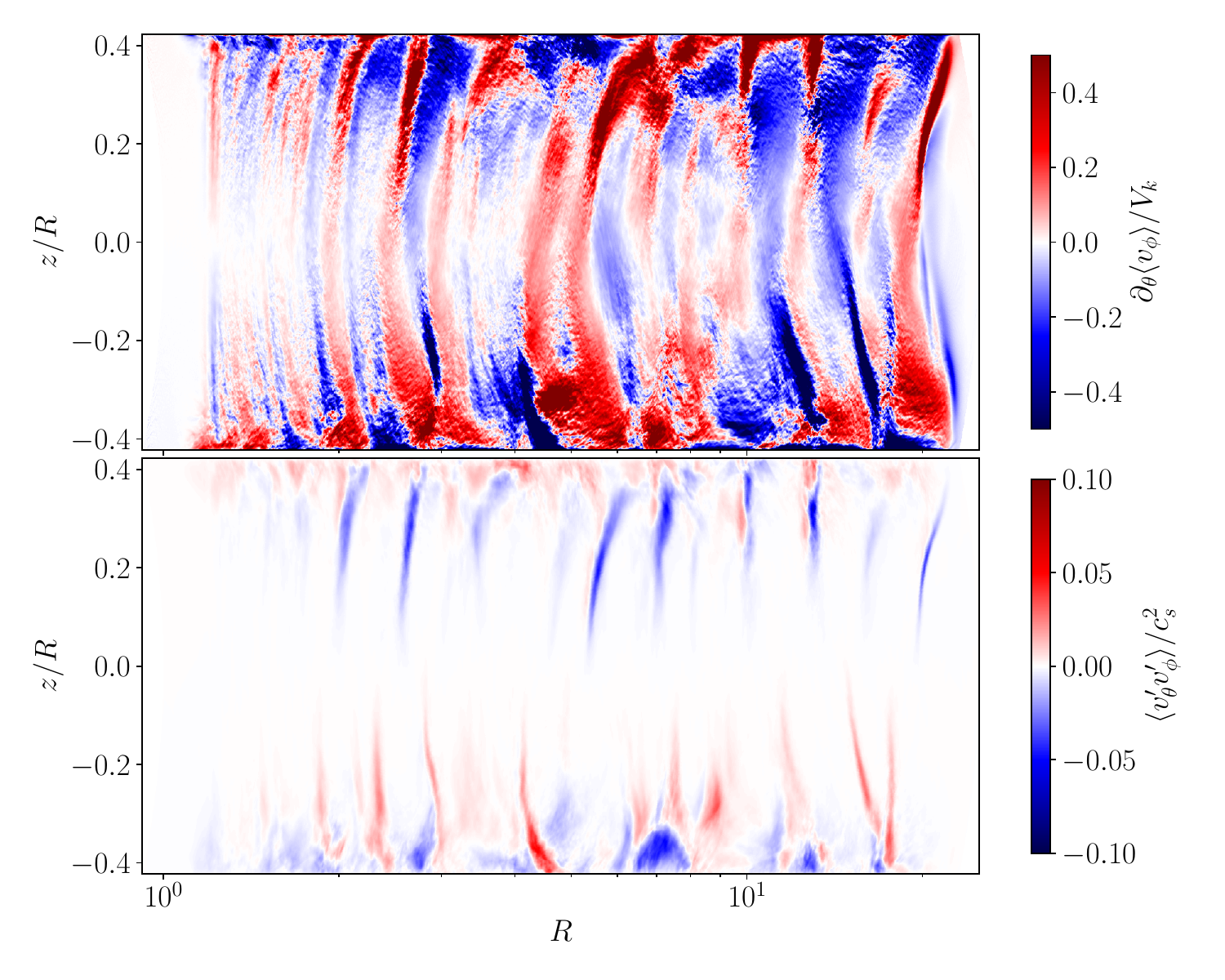}
   \caption{\label{fig:shear-stress-2D}Illustration of the correlation observed between the axisymmetric shear, $\partial \langle v_\phi\rangle/\partial \theta$ (top), and the non-axisymmetric stress tensor, $\langle \delta v_\theta\delta v_\phi\rangle$ (bottom), motivating a turbulent viscosity approach. }
   \end{figure}

Following this idea, we tested the hypothesis that the non-axisymmetric  flow acts as an effective viscosity on the VSI-unstable axisymmetric flow. Hence, we looked for correlations between the azimuthally averaged shear rate $\partial \langle v_i\rangle /\partial x_j$ and the azimuthally averaged stress tensor due to non-axisymmetric fluctuations $\langle v_i' v_j'\rangle $, where $\langle \cdot \rangle$ denotes a $\phi$-average and $ v_i'=v_i-\langle v_i\rangle$. To support this hypothesis, Fig.~\ref{fig:shear-stress-2D} compares the  $\phi-\theta$ shear and stress components, revealing that the largest turbulent stresses occur where the large-scale shear is strongest. This motivated us to look for general correlations of the form
\begin{align}
\langle v_i' v_j'\rangle = \nu_{	ijkl}\partial \langle v_l\rangle /\partial x_k,
\end{align}
where $\nu$ is a fourth-order tensor describing the `turbulent viscosity' induced by the small scales on the large axisymmetric scales. By construction, $\nu_{ijkl}$ is symmetric with respect to $i$ and $j$.

\begin{figure}
   \centering
\includegraphics[width=1.0\linewidth]{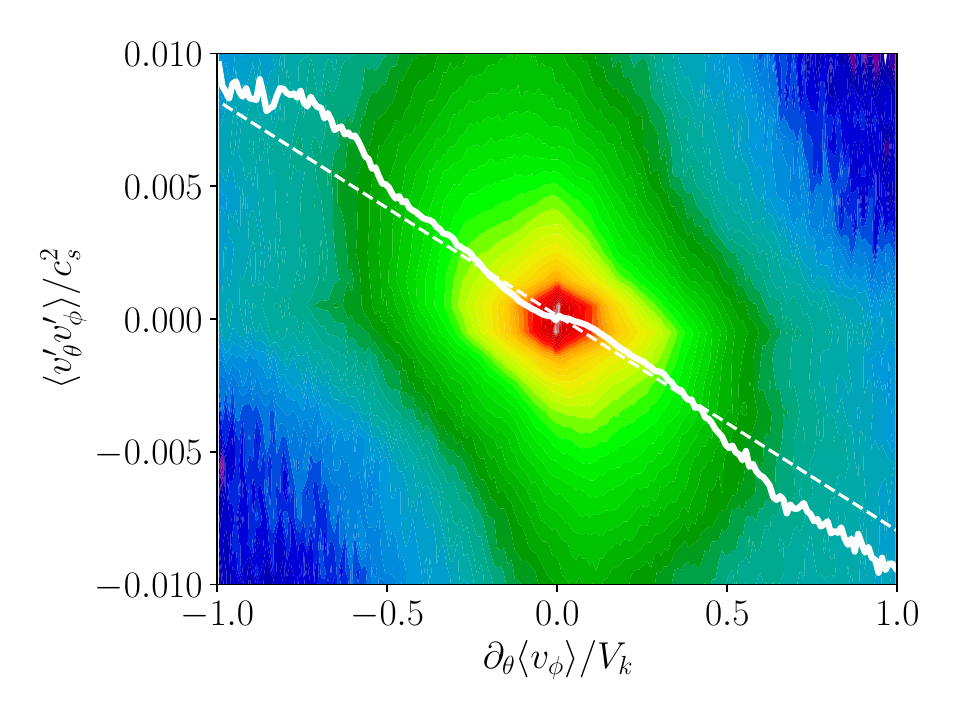}
\includegraphics[width=1.0\linewidth]{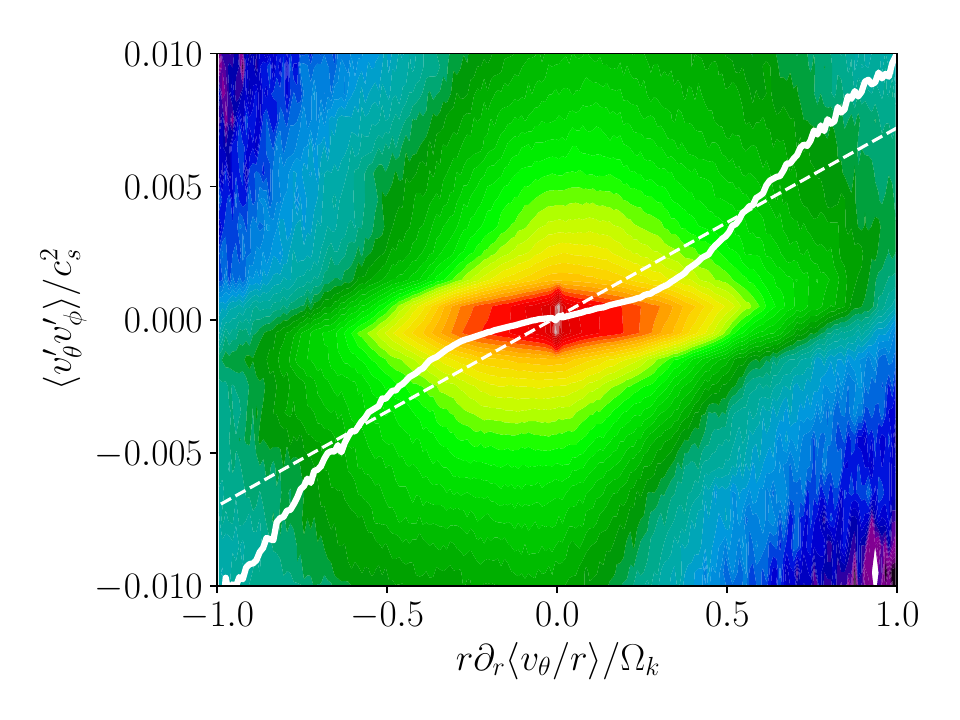}
   \caption{\label{fig:shear-stress-prob}Probability density function (PDF) of the $\theta-\phi$ (top) and $r-\theta$ (bottom) components of shear versus the $\theta-\phi$ component of the turbulent stress tensor.  The turbulent stress expectation for a given shear rate is over-plotted as a solid white line, and the dashed line shows the linear regression of the PDF with turbulent viscosities $\nu_{\theta \phi\theta \phi}=8.1\times 10^{-3} c_sH$ (top) and $\nu_{\theta \phi r\theta}=-7.1\times 10^{-3} c_sH$ (bottom). }
   \end{figure}

To quantify the relation, Fig.~\ref{fig:shear-stress-prob} (top panel) presents the probability density function (PDF) of the turbulent $\theta-\phi$ stress versus the mean vertical shear $\partial_\theta v_\phi$ computed from six 3D snapshots of run HX at $t=400,500,600,700,800$, and $900$ orbits to produce a temporal average of the stress-shear correlation. This PDF recovers the correlation previously observed in Fig.~\ref{fig:shear-stress-2D}. However, a large dispersion is clearly present. A tentative linear fit (dashed line) to the PDF yields an effective viscosity $\nu_{\theta \phi\theta \phi}=8.1\times 10^{-3} c_sH$. The stress expectation (solid line) suggests a non-linear relation, showing a slightly stronger turbulent stress response for stronger absolute shear, indicative of a larger turbulent viscosity for larger shears, which steepens as the shear increases. A similar stress response is clearly observed in the correlation between the $\theta-\phi$ stress versus the mean vertical shear $\partial_r v_\theta$ (Fig.~\ref{fig:shear-stress-prob}, bottom panel), where the fluctuation response increases non-linearly with shear.

Employing such a non-linear turbulent viscosity (i.e., one that depends on the shear) could mimic the effect of 3D small-scale turbulence in 2D large-eddy simulations of the VSI. We also note that the correlation between the $\theta-\phi$ turbulent stress and the $\partial_r v_\theta$ shear is unexpected in the context of `standard' turbulent viscosity, which typically assumes the viscous tensor to be only non-zero for the components $\nu_{ijij}$. This `odd viscosity' produces peculiar behaviours in complex flows \citep[e.g.][]{Fruchart.Scheibner.ea23} that are yet to be understood in the context of turbulent protoplanetary discs. Our findings demonstrate that the turbulent stress response is not only non-linear, but also anisotropic and non-diagonal, confirmed by examination of all possible stress-shear correlations (Appendix \ref{sec:stress}). Developing a turbulent closure scheme is beyond the scope of this work, but our results indicate how to design a turbulent-viscosity model for VSI saturation.

These results suggest that VSI saturation occurs mainly in regions of strong mean axisymmetric shear, where the turbulent response is the largest. Since the strongest shear layers are found at 2--3 scale heights (Fig.~\ref{fig:shear-stress-2D}, top), this indicates that saturation (and therefore energy deposition) probably occurs in the disc atmosphere, rather than in the disc midplane, as a naive $\alpha$-disc model would suggest.

\section{Conclusions}

In this study, we present a detailed numerical investigation of the vertical shear instability (VSI) in locally isothermal protoplanetary discs, focusing on its nonlinear saturation, wave dynamics, and turbulence in 3D. We designed the simulations to capture several large-scale wave zones and a significant portion of the turbulent cascade below the disc scale height ($H$). Our key findings are summarised below.

First, we find that bulk properties, such as angular momentum flux and kinetic energy density, converge with increasing resolution, with as few as 50 grid points per $H$ sufficient to capture the essential VSI-driven transport. This provides a useful benchmark for future large-scale simulations that aim to include VSI physics at moderate computational cost.

Second, we observe a systematic radial increase in VSI-induced turbulence and transport efficiencies within an extended inner disc region spanning over $20H$, beyond which a true self-similar regime is reached. This trend appears to be artificial, likely stemming from the imposed inner radial boundary conditions, and is consistent with boundary-driven wave-excitation mechanisms discussed in \cite{Ogilvie.Latter.ea25}. Caution is therefore warranted when interpreting radial profiles near domain boundaries or in simulations that are not sufficiently radially extended; a minimum of $r_\mathrm{out}/r_\mathrm{in}\gtrsim 5$ appears necessary to reach an asymptotic regime.

Unlike some previous studies, we do not detect long-lived zonal flows or coherent vortices in our models. This absence may be attributable to our use of a strictly locally isothermal equation of state, which tends to make the VSI more vigorous, or, alternatively, due to our higher resolutions, which might better describe the parametric instabilities that destroy vortices. Finite cooling timescales may be necessary to sustain coherent vortex structures, and further exploration of thermodynamic modelling is needed.

The corrugation-wave phenomenology described by \cite{Svanberg.Cui.ea22}persists in our 3D simulations: the dominant inertial wave frequencies are consistent with those observed in 2D, low-resolution models. However, the 3D wave fields are significantly noisier and the mechanism responsible for this frequency selection remains unclear, warranting further theoretical investigation.

The resulting turbulence exhibits strong anisotropy at large scales, consistent with the phenomenology of critically balanced rotating turbulence. At small scales, the turbulent power spectrum exhibits a distinct break with a very steep power law, echoing the findings of\cite{Manger.Klahr18}, and reflecting numerical viscosity. 

Finally, we find that small-scale, non-axisymmetric VSI structures act as an anisotropic effective viscosity on the axisymmetric component of the flow. This emergent behaviour suggests the possibility of developing axisymmetric effective models in which full 3D turbulence is incorporated via a closure scheme, potentially reducing computational costs while retaining key physical effects.

Together, these findings contribute to a growing understanding of VSI dynamics and lay the groundwork for more physically comprehensive and computationally efficient models of protoplanetary disc evolution.

\begin{acknowledgements}
We thank Mario Flock, Wladimir Lyra, Orkan Umurhan and Oliver Gressel for useful feedback on the manuscript. 
     The authors acknowledge the EuroHPC Joint Undertaking for awarding this project access to the EuroHPC supercomputer LUMI, hosted by CSC (Finland) and the LUMI consortium through a EuroHPC Extreme Scale Access call. GL acknowledges support from the European Research Council (ERC) under the European Union Horizon 2020 research and innovation program (Grant agreement No. 815559 (MHDiscs)). HNL and GIO are supported by the Science and Technology Facilities Council (STFC) through grant ST/X001113/1.
\end{acknowledgements}

\bibliographystyle{aa} 
\bibliography{VSI-bibliography} %

\begin{appendix}
 \onecolumn
\section{Full 2D kinetic energy spectra}

\begin{figure*}[h!]
   \centering
   \includegraphics[width=0.9\linewidth]{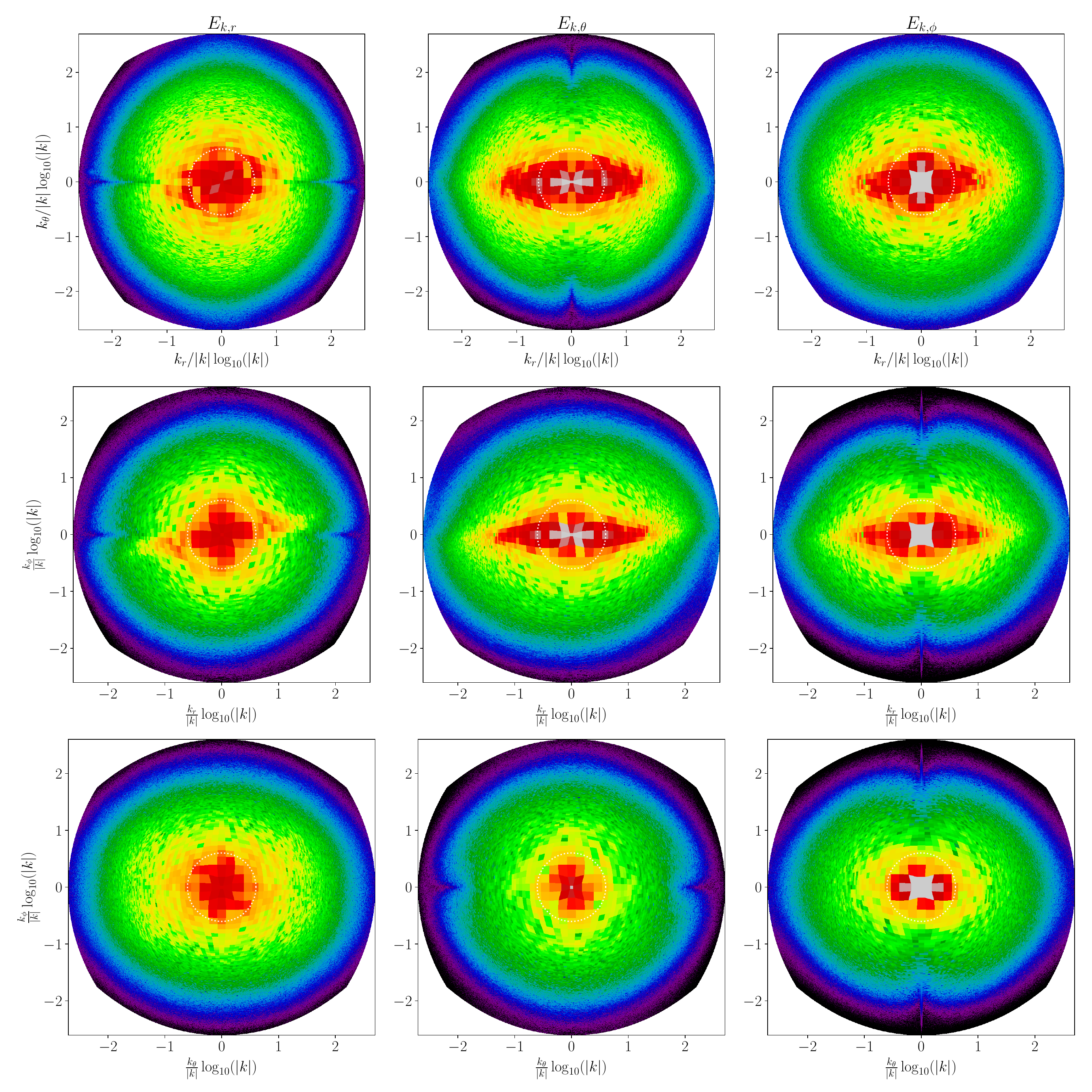}   
   \caption{\label{fig:2D-spectrum-full}2D cuts of the kinetic energy spectrum in run EX for each velocity component. The white dotted line delimits the wavenumbers $|k|=2\pi/H$.}
\end{figure*}

We show in Fig.~\ref{fig:2D-spectrum-full} the full information on the 2D spectra for each component of the velocity field.

\section{Additional shear-stress correlations}
\label{sec:stress}
\begin{figure}
   \centering
\includegraphics[width=0.5\linewidth]{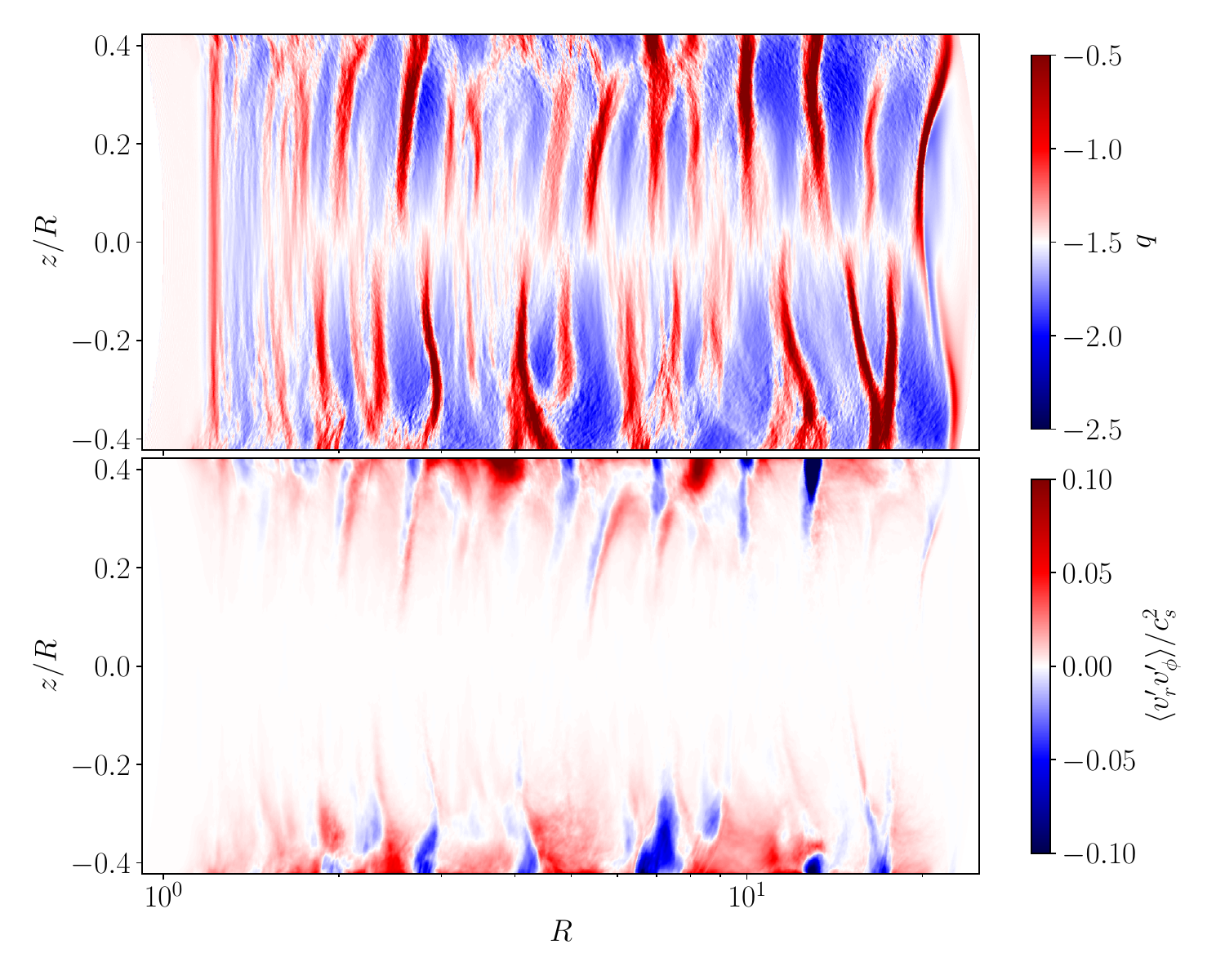}
   \caption{\label{fig:shear-stress-rphi-2D}Illustration of the correlation observed between the axisymmetric shear $q$ (top) and the non-axisymmetric stress tensor $\langle \delta v_r\delta v_\phi\rangle$ (bottom). Note that our large-scale shear includes the mean Keplerian shear which has $q=-1.5$. }
   \end{figure}

\begin{figure*}
   \centering
\includegraphics[width=0.8\linewidth]{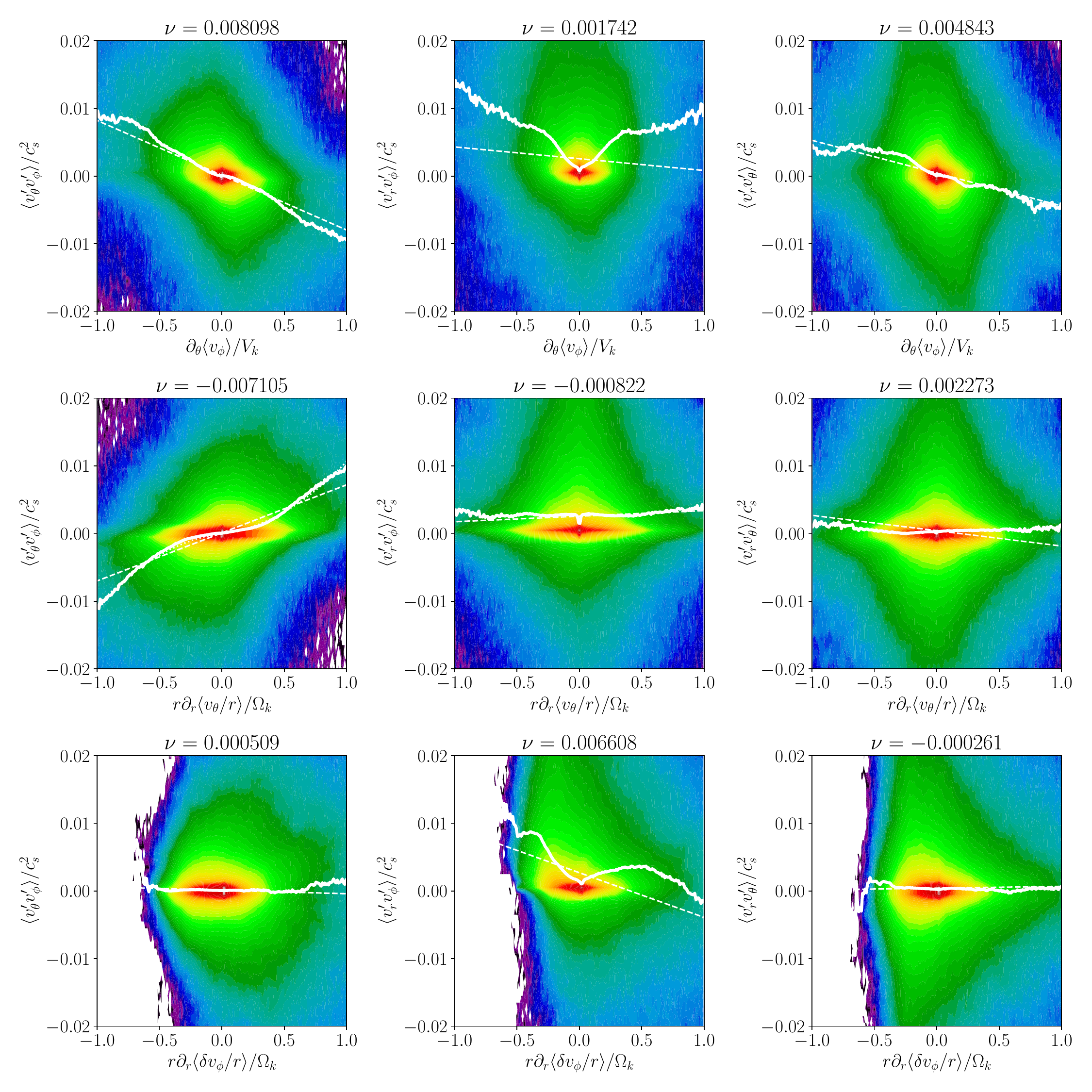}
   \caption{\label{fig:shear-stress-full-table}Probability density function (PDF) of the average shear components versus the turbulent stress tensor components. Overplotted in the solid white line is the turbulent stress expectation for a given shear rate, and in dashed line the linear regression of the PDF that yields the corresponding turbulent viscosity (given in the title of each panel).}
   \end{figure*}

We present here additional correlations between the large-scale shear and the turbulent stress from a snapshot from run EX in Fig.~\ref{fig:shear-stress-rphi-2D}. We again find a correlation "by eye" between the large-scale shear and the turbulent stress. We next move to a PDF of all of the possible combinations of stress and shear in Fig.~\ref{fig:shear-stress-full-table}. We observe several correlations that points toward a possible non-diagonal viscous tress tensor. Note however that the large-scale stresses from the axisymmetric corrugation wave are also correlated, so we can't assert for sure that the viscous stress tensor is non-diagonal from our diagnostics.

Note in addition that for PDFs looking at turbulent stresses versus $S=r\partial_r \langle \delta  v_\phi/r\rangle /\Omega_K$ (last line of Fig.~\ref{fig:shear-stress-full-table}), the system never explores the regime $S<-0.5$. The line $S=-0.5$ corresponds to marginal Rayleigh stability. We therefore confirm that axisymmetric flows never cross the Rayleigh line. 

\end{appendix}
\end{document}